\input{epsf}

\documentclass[12pt, draftclsnofoot, peerreview, onecolumn]{IEEEtran}

\makeatletter
\def\ps@headings{%
\def\@oddhead{\mbox{}\scriptsize\rightmark \hfil \thepage}%
\def\@evenhead{\scriptsize\thepage \hfil \leftmark\mbox{}}%
\def\@oddfoot{}%
\def\@evenfoot{}}
\makeatother
\usepackage{epsf}
\usepackage{graphicx}
\usepackage[cmex10]{amsmath}
\usepackage{amsthm}
\usepackage{amssymb}
\usepackage{epsfig,latexsym,amsmath,epsf,amssymb,amsfonts}
\usepackage{algorithm, algorithmic}
\usepackage{placeins}
\usepackage{url}
\usepackage{cite}
\usepackage{comment}
\usepackage{subfigure}
\usepackage{lipsum}
\usepackage{physics}
\usepackage{color}
\usepackage{float}

\usepackage{pgf,tikz}

\usepackage[top=0.75in, bottom=1in, left=0.625in, right=0.625in]{geometry}
\usepackage{bbm}

\usetikzlibrary{shapes,arrows,automata, calc}

\renewcommand{\theequation}{\arabic{equation}}
\newtheorem{Definition}{\hskip 0pt Definition}
\newtheorem{Theorem}{\hskip 0pt Theorem}
\newtheorem{Fact}{\hskip 0pt Fact}
\newtheorem{Lemma}{\hskip 0pt Lemma}
\newtheorem{Proposition}{\hskip 0pt Proposition}

\newtheorem{Remark}{\hskip 0pt Remark}
\begin{document}
\title{Decentralized Learning for Channel Allocation in IoT Networks over Unlicensed Bandwidth as a Contextual Multi-player Multi-armed Bandit Game}

\author{
  \IEEEauthorblockN{Wenbo Wang,~\IEEEmembership{Member,~IEEE,}
  Amir Leshem~\IEEEmembership{Senior Member,~IEEE},
  Dusit Niyato~\IEEEmembership{Fellow,~IEEE}
  and
  Zhu Han~\IEEEmembership{Fellow,~IEEE}\vspace*{-4mm}
  }
  \thanks{Wenbo Wang and Amir Leshem are with the Faculty of Engineering, Bar-Ilan University, Ramat-Gan, Israel 52900 (email: wangwen@bui.ac.il, leshema@biu.ac.il).}
\thanks{Dusit Niyato is with the School of Computer Science and Engineering, Nanyang Technological University, Singapore 639798
(email: dniyato@ntu.edu.sg).}
\thanks{Zhu Han is with the University of Houston, Houston, TX 77004 USA (e-mail:zhan2@uh.edu), and also with the Department of Computer Science and Engineering, Kyung Hee University, Seoul, South Korea.}
}

\maketitle
\begin{abstract}
We study a decentralized channel allocation problem in an ad-hoc Internet of Things network underlaying on the spectrum licensed to a primary cellular network. In the considered network, the impoverished channel sensing/probing capability and computational resource on the IoT devices make them difficult to acquire the detailed Channel State Information (CSI) for the shared multiple channels. In practice, the unknown patterns of the primary users' transmission activities and the time-varying CSI (e.g., due to small-scale fading or device mobility) also cause stochastic changes in the channel quality. Decentralized IoT links are thus expected to learn channel conditions online based on partial observations, while acquiring no information about the channels that they are not operating on. They also have to reach an efficient, collision-free solution of channel allocation with limited coordination. Our study maps this problem into a contextual multi-player, multi-armed bandit game, and proposes a purely decentralized, three-stage policy learning algorithm through trial-and-error. Theoretical analyses shows that the proposed scheme guarantees the IoT links to jointly converge to the social optimal channel allocation with a sub-linear (i.e., polylogarithmic) regret with respect to the operational time. Simulations demonstrate that it strikes a good balance between efficiency and network scalability when compared with the other state-of-the-art decentralized bandit algorithms.
\end{abstract}

\newpage

\section{Introduction}\label{Sec:Introduction}
The global proliferation of the Internet-connected devices has spawned a high demand for the research into supporting Internet-of-Things (IoT) communications towards next-generation wireless technologies. In novel IoT-centric use cases such as advanced metering and monitoring infrastructures for smart city/industry, the IoT networks are typically based on the co-channel deployment of the computation/power-limited Machine-Type Communication (MTC) devices~\cite{7263367,7815384} over unlicensed frequency bands. Meanwhile, these applications demand frequent transmission of small-size data either directly between MTC devices~\cite{6815897} or from these devices to IoT gateways. As a result, the unique characteristics of IoT imposes a series of challenges to the adaptation of the off-the-shelf Media Access Control (MAC) protocols in network design. Especially, the IoT networks are expected to support applications requiring a relatively high degree of efficiency and scalability, but on a basis of light-weight MAC mechanisms and minimum infrastructure, mainly due to the constraints of complexity and power resources on the devices.

In this paper, we investigate the problem of handling an anarchy group of low-complexity IoT devices for their connections underlaying over the multiple bands of a primary network. In a typical setting of heterogeneous narrow-band IoT networks, the underlaying IoT transmission can retain an efficient data rate by adopting a proper resource-block spreading factor, while causing negligible interference to the licensed cellular users by scaling the transmit power accordingly~\cite{5450024}. However, due to the impoverished resources of the low-power, light-weight IoT devices, it is impractical for them to perform simultaneous, real-time channel estimation for multiple bands with unknown, time-varying activities of the licensed users. Also, due to the limited signaling capability of the IoT devices, a pure contention-based or reservation-based channel allocation scheme (e.g., random access or coordinated access)  may not be able to meet the requirements of scalability, efficiency and reliability at the same time. For this reason, we aim to design a low-complexity, purely decentralized allocation scheme that associates the logical channels over the unlicensed bandwidth with each ad-hoc IoT link, while guaranteeing the social performance of the entire IoT network.

To achieve the two-fold goal of decentralization and social optimal performance for channel allocation, we propose a framework of decentralized strategy learning for channel association based on Multi-Player (MP) Multi-Armed Bandits (MAB). Under the proposed learning framework, the IoT devices gradually learn their link quality over each channel and then resolve the channel contention problem without explicit signaling. More specifically, we formulate the IoT devices as the players of an MAB game, and the fading channels as the stochastic arms of the MAB. To address the interference from the co-existing primary transmissions, we further extend our MP-MAB formulation by considering the underlying arm-value distribution to be non-stationary.
In particular, the instantaneous quality of an IoT link established over a primary channel\footnote{In what follows, we use the pairs of terms ``channel'' and ``arms'', and ``links'' and ``players'' interchangeably.} is determined by not only the stochastic channel state, but also the context of primary transmissions that the coexisting licensed users happen to operate upon. This leads to a multi-player extension of the contextual MAB~\cite{zhou2015survey}, where the reward of each arm for every player is jointly determined by the context (i.e., the radio environment information) and the players' actions. The goal of investigating this contextual MP-MAB is to find optimal policies mapping from the random samples of context-reward pairs over the licensed channels to a sequence of actions of channel association, to maximize the accumulated sum of achievable throughput (i.e., received reward) by the IoT links along the time horizon.

The rest of the paper is organized as follows: Section~\ref{sec_related_work} discusses the related works in the recent literature. Section~\ref{sec_model} mathematically transform the considered channel-allocation problem into a contextual MP-MAB game, based on which we propose the purely decentralized social-optimal policy learning algorithm in Section~\ref{sec_bandit_algorithm}. The efficiency of the proposed algorithm is mathematically analyzed in Sections~\ref{sec_game_analysis} and~\ref{sec_unobservable}, where Section~\ref{sec_game_analysis} provides the theoretical bound on the regret of the proposed algorithm for contextual MP-MAB, and Sections~\ref{sec_unobservable} analyzes the efficiency of the proposed algorithm when the contexts are not observable to the players. Section~\ref{sec_simulation} provides a series of experiment/simulation results regarding the proposed algorithm for heterogeneous IoT networks over unlicensed spectrum. Finally, Section~\ref{sec_simulation} concludes the paper.

\section{Related Work}
\label{sec_related_work}
\subsection{Channel Access Mechanisms for IoT Networks}\label{review_NB_IoT}
The reservation-based MAC protocols, such as the canonical F/T/C-DMA or OFDMA schemes, are able to achieve deterministic Quality of Service (QoS). However, the need of coordination over a dedicated feedback/control channel by the Access Points (APs)  (e.g.,~\cite{8664581}) limits their deployment to those scenarios in a cellular infrastructure. Comparatively, the contention-based protocols, e.g., ALOHA and CSMA/CA, are able to support larger scale M2M networks but face the issue of providing only opportunistic QoS guarantee. As a result, the hybrid MAC scheme is studied by a number of works in the literature~\cite{6762845,hegazy2017efficient} when a centralized coordinator (e.g., AP) is available in the IoT network. The hybrid MAC schemes are featured by the aggregation of the contention-based and the reservation-based protocols. They allow the network to guarantee fairness among contending users with ALOHA/CSMA-like schemes. The rate efficiency is then provided with pre-allocated orthogonal resources (e.g., time slots or sub-carriers) to a selected group of devices that win the resource request contention.

When APs (namely, infrastructure) are non-existent for the IoT networks over shared spectrum, random-access based on contentions becomes more suitable than the reservation-based schemes. In the scenario of multi-channel association with a dedicated common control channel, channel-contention resolutions based on the RTS/CTS dialog over the control channel are designed for devices equipped with multiple antennas/sensors~\cite{7843917}. Alternatively, decentralized channel swapping mechanisms are proposed for the cases in which no coordination channel is accessible for time-synchronized nodes~\cite{7360239,8792108} or for nodes even without global synchronization~\cite{10.1145/2107502.2107533}. Usually, nodes over each channel are assumed to be fully connected to avoid the hidden terminal problem. In different studies, an operational phase of broadcasting beacon packets over randomly selected channels is commonly adopted to either determine the level of congestion~\cite{7360239} or to locate free bands~\cite{8792108,10.1145/2107502.2107533} for collision avoidance. Particular mechanisms such as master node (known as SYNC node in~\cite{7360239}) election are proposed to designate the IDs of nodes and channels that swapping/hopping is allowed for, in order to achieve a convergent solution among the decentralized devices~\cite{7360239,8792108}.

However, most contention/swapping-based decentralized MAC schemes in the literature prioritize non-colliding allocation over social-optimal network performance. Another obstacle for designing an efficient and decentralized IoT MAC scheme over multi-channels lies in the lack of CSI in an unknown time-varying wireless environment. As a result, the demand for efficient decentralized MAC schemes in IoT networks inspires the adoption of distributed stochastic learning algorithms, which range from decentralized stochastic learning automata in repeated channel allocation games~\cite{7307215} to channel allocation in a framework of MP-MAB based on distributed auction~\cite{7732984} or hopping~\cite{tibrewal2019multiplayer, 8440092} with different levels of message exchanges.

\subsection{MP-MAB for Resource Allocation in Wireless Networks}\label{review_MPMAB}
In wireless networking, the MAB-based formulation was first introduced for the single-user-multi-channel selection problem in Cognitive Radio Networks (CRNs), where channel states are stochastic and not fully observable due to the unknown activities of the primary user~\cite{4723352, 7962233}. In the single-player scenario, the player's goal is to maximize the expected accumulated reward, namely, the achievable transmit rate in the long run. When the pulled arm yields i.i.d. random rewards following a stationary but unknown distribution, such a distribution can be learned from repeated plays for abstracting the unknown wireless environment, i.e., the quality of each orthogonal channel~\cite{LAI19854}. Unlike supervised learning, the value of each arm in the MAB is not known in advance, and the player is only able to observe the value of the pulled arm, one at each time. Therefore, it is necessary to infer the best arm-values from such historical partial observation through trial and error. Essentially, the solution to this well-know problem is about striking a trade-off between policy exploitation and exploration. Namely, the player needs to properly choose whether to gain the myopic optimal reward, or to further improve its arm-value estimation in order to avoid choosing a sub-optimal arm in the long run. The former goal is achieved by selecting the best arm/channel according to the available observation record, while the latter is achieved through proper policy exploration.

It is natural to extend the problem formulation from single-player MAB to the case of MP-MAB~\cite{5535151}, especially in the multi-link context of CRNs, which have to frequently deal with unknown stochastic channels due to the unpredictable activities of primary users~\cite{6151769,8792155}. With decentralized and simultaneous arm selection, collisions have to be handled when more than one player choose the same arm. At each round of play in the MP-MAB, every player chooses one arm to pull according to its own observation history of arm-value feedback, while a certain level of coordination (i.e., messaging between devices) may be allowed based on different assumptions of information exchange capabilities~\cite{7765076, besson2017multi, 7867071} for collision resolution. The rewards of the same arm observed by different players are frequently assumed to be drawn from different and unknown i.i.d. distributions (e.g.,~\cite{8792108}), which reflects the independent pathloss and shadowing properties of different user links over the same channel. With such a multi-player formulation, a repeated game of heterogeneous players evaluating player-dependent rewards over the candidate channels can be developed (e.g.,~\cite{8792155,7398139}).

\subsection{Contribution}\label{contribution}
Compared with the existing studies in the literature, our research further extends the expressiveness of the MP-MAB formulation with a new dimension of freedom brought by the environmental context information~\cite{dudik2011efficient}. In the considered scenario of coexisting IoT network operation over spectrum licensed to primary users, we employ a discrete set of contexts to quantitatively reflect the interference caused to the ad-hoc IoT links by the primary transmissions. Such formulation allows us to further address the randomness of the network environment, which can be widely observed in different wireless networking protocol layers such as user hand-off between cells in a CRN~\cite{5740912} and multi-task execution with different levels of QoS~\cite{7842265}. However, the introduction of the discrete context requires the MP-MAB algorithm designer to reconsider the regret propagation from the very beginning of the learning process as well as the collision avoidance mechanism among the decentralized links. Bearing these challenges in mind, the main contributions of this paper are summarized as follows:
\begin{itemize}
  \item We model the dynamic channel allocation problem in an ad-hoc IoT network over shared spectrum as a multi-player contextual MAB. Especially, we address the problem of unknown stochasticity in both channel states and non-controllable activities of the underlying licensed users at the same time. We propose a novel decentralized online learning algorithm, which achieves social-optimal channel allocation with no need of the a-priori knowledge about channel statistics and radio context evolution.
  \item We study a generalized scenario where IoT links observe heterogeneous achievable rates over the same channel in the same radio context (e.g.,  due to different distances from the primary transmitter). Extended from a typical exploration-exploitation framework for single-player contextual MAB~\cite{langford2008epoch}, theoretical analysis is provided regarding the convergence property and the network performance, under the framework of efficient pure Nash Equilibrium (NE) selection with log-linear learning~\cite{PRADELSKI2012882}.
  \item Theoretically, we show that the proposed algorithm achieves polynomial logarithmic regret over time and can handle a large number of discrete contexts. Our simulation experiments demonstrates this by comparing the proposed algorithm with a number of state-of-the-art MP-MAB algorithms.
\end{itemize}

\section{Problem Formulation}
\label{sec_model}
\subsection{Network Model}\label{sec_network_model}
\begin{figure}[!t]
	\centering
	\includegraphics[width=.85\textwidth]{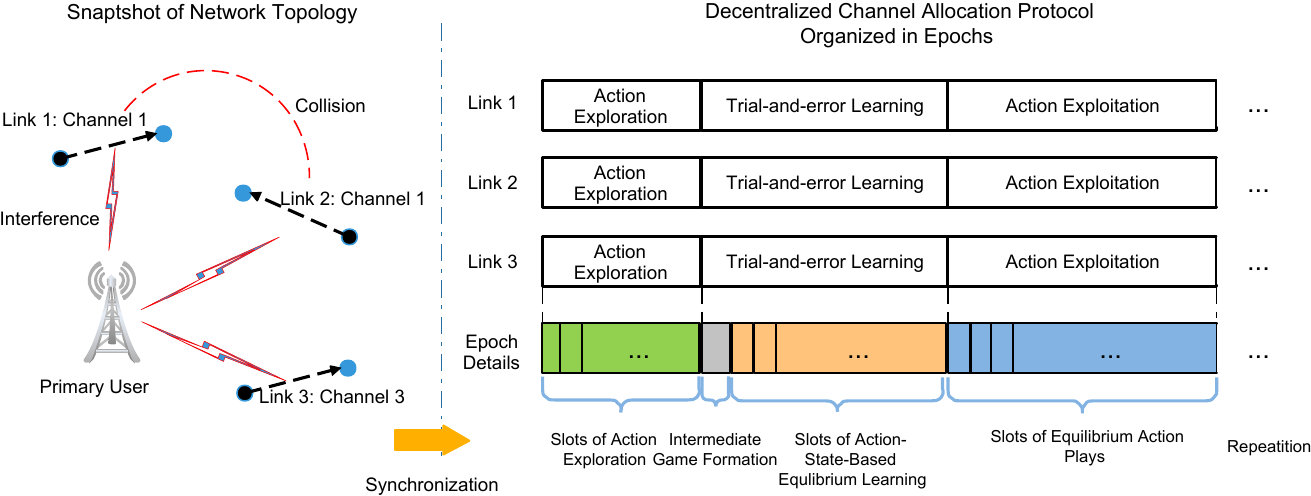}\vspace{-5mm}
	\caption{Topology of a fixed-size cognitive ad-hoc IoT network and the three phases of decentralized policy learning in synchronized time slots.}
  \label{fig_epoch_tne}
\end{figure}

We consider $M$ ad-hoc IoT links attempting to access $L$ ($L\!\ge\! M$) uncorrelated unlicensed channels\footnote{We assume that the MTC transmissions can be delayed or advanced to avoid overwhelming the available number of bands. Otherwise, all channels are occupied when $L< M$, and collision avoidance techniques are thus needed in addition to bandit-based learning.} in the underlay mode. Each link independently chooses a channel to transmit over, and each channel supports no more than one link at the same time. During the network operation, the primary users cause a random level of interference on the channels. The received signal of link $m$ over channel $l$ can be expressed as
\begin{equation}
  \label{eq_rx_signal}
  y^l_m(t)=h^l_m(t)s_m(t) + h^l_{p,m}(t)s_p(t) + n_l(t),
\end{equation}
where $h^l_m(t)$ and $h^l_{p,m}(t)$ are the coefficients of channel $l$ for the IoT link $m$ and its received interference from the primary transmission, each of which is sampled from an unknown i.i.d. probability distribution over time, determined by both the unknown stochastic device-mobility patterns and small-scale fading. $s_m(t)$ and $s_p(t)$ are the transmit signals of the IoT link and the primary link, respectively. $n_l(t)$ is the Additive White Gaussian Noise (AWGN) with an unknown variance $\sigma_l^2$. We consider that the IoT links operate in a synchronous time-slotted manner, and the operating slot is set to be of the same timescale as the coherence time of the fading channel. We adopt the mild assumption that the primary interference dominates the perceivable interference plus noise at the IoT receivers. At the beginning of a time slot, the IoT devices are able to sense the instantaneous, discrete levels of the licensed transmission power over the spectrum either by employing a simple energy detector or through the feedback of the primary base station.
For example, without the presence of the IoT transmission, the finite discrete power level of the primary user over channel $l$ is measured by $Y^l_m=\sum_{i=1}^{N_s}\Vert{y^l_m}(i)\Vert^2$ as $x_l$, where $N_s$ is the number of samples collected during the sensing sub-slot.
However, the IoT links do not know the stochastic activity (e.g., power-selection) patterns of the licensed users. The network structure is shown in Figure~\ref{fig_epoch_tne}.
\label{sec_bandit_algorithm}

Due to the limited signal processing capability, an IoT device is only able to measure the link QoS in best-effort bitrate over its selected channel at the end of a time slot. When no collision happens during time slot $t$, this can be upper-bounded by the achievable throughput, which is proportional to:
\begin{equation}
  \label{eq_rx_throughput}
  r^l_m(t)= \log_2\left(1+\frac{{\Vert h^l_m(t) \Vert}^2P_m}{\Vert {h^l_{p,m}(t) \Vert }^2P_p+\sigma_l^2}\right),
\end{equation}
where $P_m$ and $P_p$ are the transmit power of IoT link $m$ and the primary link, respectively. Otherwise, link $m$ will fail to deliver any data on the colliding channel and observe zero throughput. The IoT transmitter-receiver pairs are expected to (implicitly) learn device-mobility and the interference patterns from only the QoS measurement of its selected channels over time. In addition, they have to infer the channel-selection policies of the other links without inter-link messaging, not only to avoid collisions with other devices but also to learn its policies toward the social optimal allocation.

\subsection{Decentralized Channel Allocation as a Contextual MP-MAB}\label{sec_MP-MAB_model}
Based on the network model described in Section~\ref{sec_network_model}, we formulate the decentralized channel selection problem as an $M$-player, $L$-arm contextual bandit game.
Following the discussion above, let ${x}$ denote the context variable describing the power level of the licensed user over the bandwidth. From (\ref{eq_rx_signal}), we note that with the licensed user evenly distributing its power over the entire spectrum, the primary interference to an IoT link is homogeneous over each logic channel\footnote{For heterogeneous power levels over different channels obtained through the feedback of the primary base station, the context can be represented by a vector $\mathbf{x}=[x_1,\ldots, x_L]^{\textrm{T}}$, which does not affect our discussion in this paper.}.
However, It varies for different IoT receivers due to their different distances from the primary transmitter. We consider that $x$ is discretized into a finite context space $\mathcal{X}$ with the cardinality $|\mathcal{X}|=X$. The context evolution of the primary transmission power level is independent of the IoT links' choices of channel association, and follows an unknown stationary random process. When no collision over the channels occurs, the players receive rewards from their selected arms by measuring the normalized QoS feedbacks according to (\ref{eq_rx_signal}) over that channel. The players repeatedly play the bandit game by simultaneously selecting the channels to operate on without any inter-link coordination. Our aim of designing the bandit-based learning mechanism is to maximize the sum of best-effort transmit rates of all the players, which is accumulated over a finite but unknown time horizon $T$. Mathematically, we abstract the contextual MP-MAB game for channel allocation in the ad-hoc IoT network as follows.
\begin{Definition}[Contextual MP-MAB]
  \label{def_contextual_bandit}
  In an $M$-player, $L$-arm contextual bandit game, there is a distribution $D_m$ for each player $m$ ($1\!\le\! m\!\le\! M$) over the context and arm-values $(x, r_{m,1}, \ldots, r_{m,L})$. $r_{m,l}\!\in\![0,1]$ is the normalized reward of player $m$ on arm $l\!\in\!\{1,\ldots, L\}$. During the repeated play, $x^t$ is drawn and revealed from the independent, unknown context distribution before round $t$, and the arm rewards for each player $m$ are sampled from $D_m$. After the players take simultaneous actions $\mathbf{a}^t\!=\![a^t_1,\ldots, a^t_M]^{\top}$, player $m$ receives a reward $r^t_{m, a^t_m}\!\in\![0,1]$ when no collision occurs over its selected arm $l\!=\!a^t_m$. Otherwise, it receives 0.
\end{Definition}

Let $\mathbf{v}^t=[v^t_1, \ldots, v^t_M]^{\top}$ denote the vector of instantaneous rewards received by the $M$ players in round $t$. Then, by taking into account the collision of players over a pulled arm, we obtain the reward of player $m$, $\forall m\in\{1,\ldots, M\}$ as
\begin{equation}
  \label{label_instantaneous_reward}
  v^t_{m}(\mathbf{a}^t) = r_{m, a^t_m}\mathbbm{1}\left(\sum_{i=1}^{M} \mathbbm{1}(a^t_{i}, a^t_{m}), 1\right),
\end{equation}
where $\mathbbm{1}(a,b)$ is the indicator function with $\mathbbm{1}(a,b)\!=\!1$ if $a\!=\!b$ and $\mathbbm{1}(a,b)\!=\!0$ otherwise.
$r_{m, a^t_m}$ is the normalized achievable throughput of link $m$ based on (\ref{eq_rx_throughput}) when no collision occurs over the selected channel according to action $a^t_m$.
Let $\mathbf{r}^t=[r^t_{m, l}]^{\top}_{1\le m\le M, 1\le l\le L}$ denote the vector of the players' arm values with respect to context $x$, and $D$ be the arbitrary distribution of the pair $(x, \mathbf{r})$.
Then, we aim to develop an algorithm that determines the joint policy $\pi(x): \mathcal{X}\rightarrow \{1,\ldots, L\}^M$ to maximize the social utility, $\sum_{t=1}^TE_{(x^t, \mathbf{r}^t)\sim D}\left\{\sum_{m=1}^M{v}_m^t(\pi(x^t))\right\}$, i.e., the expected accumulated reward of all the IoT links.
To help examining the performance of our algorithm, we introduce the concept of regret as follows.
\begin{Definition}[Regret for Observable Contexts]
  \label{def_regret}
  Let $V(\pi)=E_{(x^t, \mathbf{r}^t)\sim D}\left\{\sum_{m=1}^M{v}_m^t(\pi(x^t))\right\}$ denote the expected reward of a joint policy $\pi$. Let $Z^T=\{(x^1, \mathbf{r}^1),\ldots, (x^T, \mathbf{r}^T)\}$ denote a series of $T$ context-value pairs drawn from the distribution $D$. Then, for an algorithm $\mathcal{B}$ that generates a corresponding series of policies $\mathcal{B}=\{\tilde\pi^1,\ldots, \tilde\pi^T\}$, the expected regret of $\mathcal{B}$ with respect to a policy $\pi$ is
  \begin{equation}
    \label{label_regret_wrt_pi}
    \Delta R(\mathcal{B}, \pi, T) = TV(\pi) - E_{Z^T\sim D}\left[\sum_{t=1}^T\sum_{m=1}^M v^t_{m}(\tilde\pi^t_{m})\right].
  \end{equation}
  The regret of algorithm $\mathcal{B}$ with respect to policy space $\Pi$ is
  \begin{equation}
    \label{label_regret}
    \Delta R(\mathcal{B}, \Pi, T) = \sup_{\pi\in\Pi} TV(\pi) - E_{Z^T\sim D}\left[\sum_{t=1}^T\sum_{m=1}^M v^t_{m}(\tilde\pi^t_{m})\right].
  \end{equation}
\end{Definition}

An efficient decentralized policy-learning algorithm $\mathcal{B}$ has to achieve a sublinear regret $\Delta R(\mathcal{B}, \Pi, T)$ in $T$, namely, $\lim\limits_{T\rightarrow \infty} \Delta R(\mathcal{B}, \Pi, T)/T=0$. Due to the partial observability that rewards are only revealed for the pulled arms, an efficiently algorithm needs to form the unbiased estimation of arm values in order to learn the accurate matching between the arms and the players. Furthermore, a purely decentralized algorithm needs to avoid requesting excessive exchange of action information among players. Thus, learning the optimal arm-allocation schemes solely based on their local information is preferred.

\section{Epoch-Based Policy Learning Algorithm}
Since the number of playing rounds $T$ is not known in advance, we divide the process of decentralized learning $\mathcal{B}$ in repeated plays into epochs/mini-batches, each of which contains three explicit phases of policy exploration, optimal-policy learning and policy exploitation. The right-hand side of Figure~\ref{fig_epoch_tne} shows the structure of one epoch of plays from the perspective of synchronized IoT devices. During the exploration phase, the players independently try different arms uniformly at random in order to estimate the mean value of the payoff obtained on each arm when no collision occurs. Consequently, with the observation accumulated in the exploration phase, the players adopt a purely decentralized learning scheme through trial-and-error from~\cite{PRADELSKI2012882} to learn the optimal arm association. This is achieved through distributively searching the social optimal equilibria of a group of intermediate non-cooperative games, which are constructed based on the arm-value estimation obtained in the exploration phase for different contexts. In the exploitation phase, the players stick to the policies derived from the policy-learning phase for multiple rounds. Intuitively, the estimation of the expected reward for each arm-player pair may introduce errors, and the arm allocation learned in the policy learning phase may be sub-optimal as well. As a result, the main goal of our study is to analyze the error propagation from the first two phases and determine the bound of the regret of the entire learning process subsequently.

The policy-learning algorithm in the mini-batch framework shown in Figure~\ref{fig_epoch_tne} is formally presented in Algorithm~\ref{alg_bandit}. In the $k$-th epoch, the number of rounds needed for a player in the phases of exploration, trial-and-error learning and exploitation are functions of the epoch number, i.e., $f(k)$, $g(k)$ and $h(k)$, respectively. In the exploration phase (Lines 3-9 in Algorithm~\ref{alg_bandit}), a single player learns independently its expected payoff over each arm by randomly selecting its actions. In the trial-and-error learning phase, a group of intermediate non-cooperative games are formulated based on the arm-values estimated in the exploration phase for each context (see Lines 11-17 in Algorithm~\ref{alg_bandit}). The optimal player-arm matching scheme in each context is learned in a purely decentralized manner with respect to the intermediate game (Lines 18-21 in Algorithm~\ref{alg_bandit}). More specifically, instead of updating the policies according to the immediate feedback of the random arm-values in each round, the players learn their policies in the intermediate game by fixing the value of each arm as the estimated rewards obtained from the previous exploration phase. Following Algorithm~\ref{alg_trial_and_error}, the optimal policies are learned as the efficient Nash Equilibria (NE) of the intermediate games. For a context $x\in\mathcal{X}$ appearing in epoch $k$, we use the vector of the estimated expected arm-values $[\mu^k_{m,l}(x)]^{\top}_{1\le m\le M, 1\le l\le L}$ in (\ref{eq_update_mean}) to construct an intermediate $M$-player non-cooperative game $\mathcal{G}(x)$ as follows.
\begin{Definition}[Intermediate Non-cooperative Game]
  \label{def_inter_game}
  The intermediate game $\mathcal{G}(x)$ at the $k$-th epoch for context $x$ can be expressed in a three-tuple: $\mathcal{G}(x)=\langle\mathcal{M}, \times\mathcal{A}_m, \{u^x_m\}_{m\in\mathcal{M}}\rangle$, where $\mathcal{M}=\{1,\ldots, M\}$ is the set of players, $\mathcal{A}_m=\{1,\ldots, L\}$ is player $m$'s action set corresponding to the candidate arms, and $u^x_m=u^x_m(\mathbf{a})$ is the payoff of player $m\in\mathcal{M}$ under a joint action $\mathbf{a}=[a_1,\ldots, a_M]^{\top}$:
  \begin{equation}
    \label{eq_payoff_inter_game}
    u^x_m(\mathbf{a}) = \mu^{k}_{m,a_m}(x)\mathbbm{1}\left(\sum_{i=1}^M\mathbbm{1}(a_i, a_m), 1\right),
  \end{equation}
  where $\mu^{k}_{m,a_m}(x)$ is the expected reward of arm $l=a_m$ that player $m$ estimates in the $k$-th exploration phase, derived following (\ref{eq_update_mean}).
\end{Definition}
\begin{algorithm}[!t]
    \begin{small}
 \caption{Policy learning at player $m$ in the contextual multi-player bandit game.}
 \begin{algorithmic}[1]
 \REQUIRE
 Set $\mathcal{W}_m=\{\}$ and $u^k_{m,l}(x)=0$ $\forall l \in\{1,\ldots, L\}$ and $\forall x\in\mathcal{X}$. Choose $\epsilon\in[0,1]$\vspace{-1mm}
 \FOR {Epoch $k=1, \ldots, k_T$}\vspace{-1mm}
  \STATE \textit{Exploration phase:}\vspace{-1mm}
  \FOR {$t = 1, \ldots, f(k)$}\vspace{-1mm}
  \STATE Sample an arm $a^t_{m}\in\{1,\ldots,L\}$ uniformly at random and observe the feedback $(x^t, a^t_{m}, v^t_{m}(\mathbf{a}^t))$\vspace{-1mm}
  \IF {$v^t_{m}(\mathbf{a}^t)\ne0$}\vspace{-1mm}
    \STATE $\mathcal{W}_m\leftarrow \mathcal{W}_m\cup \{(x^t, a^t_{m}, v^t_{m}(\mathbf{a}^t))\}$\vspace{-1mm}
    \STATE Estimate the expected value of arm $l=a^t_{m}$ at $x^t$:\vspace{-2mm}
    \begin{equation}
      \label{eq_update_mean}
      \mu^k_{m,l}(x^t)\leftarrow \frac{\sum_{(x, a_{m}, v_{m})\in\mathcal{W}_m}v^t_{m}\mathbbm{1}(a_m, l)\mathbbm{1}(x, x^t)} {\sum_{(x, a_{m}, v_{m})\in\mathcal{W}_m}\mathbbm{1}(a_m, l) \mathbbm{1}(x, x^t)}\vspace{-2mm}
    \end{equation}
  \ENDIF\vspace{-1mm}
  \ENDFOR\vspace{-1mm}
  \STATE \textit{Trial-and-error learning phase:}\vspace{-1mm}
  \STATE $\forall x\!\in\!\mathcal{X}$, construct game $\mathcal{G}(x)$ as in Definition~\ref{def_inter_game}. Namely, $\forall m\!\in\!\mathcal{M}, l\!\in\!\mathcal{A}_m$, fix the perturbed arm-value as $\tilde{\mu}^k_{m,l}(x)=\mu^k_{m,l}(x)+\xi_{m,l}(x)/k$,
  where $\xi_{m,l}(x)$ is randomly sampled over $[-\xi, \xi]$ with $0<\xi<1$\vspace{-1mm}
  \IF {$k=1$}\vspace{-1mm}
  \STATE $\forall x\!\in\!\mathcal{X}$, set the auxiliary state at $t\!=\!0$ in (\ref{label_content_state}) as
  $z^0_{m}(x)\!=\!(o^0_m(x)\!=\!D, \overline{a}^0_m(x), \overline{u}^0_{m}(x)\!=\!0)$ with a random action $\overline{a}^0_m(x)$\vspace{-1mm}
  \ELSE\vspace{-1mm}
    \STATE $\forall x\!\in\!\mathcal{X}$, initialize $z^0_{m}(x)$ with the exploitation policy in epoch $k-1$ as $z^0_{m}(x)\!=\!(o^0_m(x)\!=\!C, {a}^{*,k-1}_m(x), \overline{u}^0_{m}(x)\!=\!0)$\vspace{-1mm}
  \ENDIF\vspace{-1mm}
  \STATE $\forall {a}_m\!\in\!\mathcal{A}_m$, set the count of times for getting a content mood with ${a}_m$ as $\nu^k_{m,x}({a}_m)=0$\vspace{-1mm}
  \FOR {$t=1 \ldots, g(k)$}\vspace{-1mm}
  \STATE Update $a^t_m$ and $z^t_{m}(x^t)$ according to Algorithm~\ref{alg_trial_and_error} for $x^t$ based on $\tilde{\mu}^k_{m,l}(x^t)$, and $\forall x'\ne x^t$, set $z^t_{m}(x')\leftarrow z^{t-1}_{m}(x')$\vspace{-1mm}
  \STATE Update the frequency of visits to the content states aligned with benchmark values as\vspace{-2mm}
  \begin{equation}
    \label{eq_frequency_visit}
    \nu_{m, x^t}^k(a^t_m)\leftarrow \nu_{m,x^t}^k(a^t_m) + \mathbbm{1}(o^t_m, C)\mathbbm{1}(u^t_m(x^t), \overline{u}^k_{m}(x^t)),\vspace{-2mm}
  \end{equation}
  where $u^t_m(x^t)$ is the observed payoff in the intermediate game $\mathcal{G}(x^t)$ by player $m$ according to (\ref{eq_payoff_inter_game})
  \ENDFOR\vspace{-1mm}
  \STATE \textit{Exploitation phase:}\vspace{-1mm}
  \FOR {$t=1, \ldots, h(k)$}\vspace{-1mm}
  \STATE For $x^t$, play $a^{*,k}_m$ with the maximum number of state visits according to $\nu_{m, x^t}^k(l), \forall l$:\vspace{-2mm}
  \begin{equation}
    \label{eq_maximum_count}
    a^{*,k}_m(x^t)=\arg\max_{1\le l\le L}\nu_{m,x^t}^k(l)\vspace{-2mm}
  \end{equation}
  \ENDFOR\vspace{-1mm}
 \ENDFOR\vspace{-1mm}
 \end{algorithmic}
 \label{alg_bandit}
\end{small}
\end{algorithm}

The design of the intermediate games in Definition~\ref{def_inter_game} is based on the presumption that the most efficient equilibria of the constructed intermediate games for each $x$ coincide with the social-optimal policies of the MP-MAB game. The detailed discussion on the validity of this presumption is inspired by the analysis of log-linear learning in~\cite{PRADELSKI2012882} and will be presented in  Section~\ref{sec_game_analysis}. To develop a purely decentralized policy-learning scheme in Algorithm~\ref{alg_trial_and_error} for obtaining the social-optimal equilibrium of the intermediate game $\mathcal{G}(x)$, we introduce the auxiliary state of player $m$ regarding context $x$ at time slot $t$ from~\cite{PRADELSKI2012882} as follows:
\begin{equation}
  \label{label_content_state}
  z^t_{m}(x)=\left(o^t_{m}(x), \overline{a}^t_{m}(x), \overline{u}^t_{m}(x)\right),
\end{equation}
where $o^t_{m}(x)\in\{C, H, W, D\}$ indicates the \emph{moods of player} $m$: content ($C$), hopeful ($H$), watchful ($W$) and discontent ($D$). $\overline{a}^t_{m}(x)$ represents the \emph{benchmark action} and $\overline{u}^t_{m}(x)$ represents the \emph{benchmark payoff} adopted by player $m$ in round $t$, respectively. For simplicity, we omit $x$ in the expressions for the same game and define the following transition map of a finite behavior state machine for each type of players:
\begin{itemize}
  \item {\bf{A content player}} updates its action as $a^t_{m}\in\mathcal{A}_m$ with a probability:
  \begin{equation}
    \label{label_cotent_prob}
    p_{m}({a^t_{m}})=\left\{
    \begin{array}{ll}
      \displaystyle\frac{\epsilon}{L-1}, & a^t_{m}\ne \overline{a}^{t-1}_{m},\\
      1-\epsilon, & a^T_{m} = \overline{a}^{t-1}_{m}.
    \end{array}\right.
  \end{equation}
  \item {\bf{A hopeful player}} or {\bf{a watchful player}} always plays the previous benchmark action, i.e., $a^t_{m}\leftarrow\overline{a}^{t-1}_{m}$.
  \item {\bf{A discontent player}} selects a new action uniformly at random, namely, $\forall a^t_{m}\in\mathcal{A}_m$, $p_{m}({a^t_{m}})=1/L$.
\end{itemize}

\begin{algorithm}[!t]
    \begin{small}
 \caption{A single round of state transition by player $m$ in $\mathcal{G}(x)$ at the $k$-th epoch.}
 \begin{algorithmic}[1]
 \REQUIRE {$\forall l\in\{1,\ldots, L\}$, retrieve the fixed, perturbed arm-value in the current context $x$ as $\tilde{\mu}^k_{m,l}(x)$}
 \STATE Select $a^t_{m}$ according to state $z^{t-1}_m(x)$ of the player, and observe $u_m^t(x)=u_m^x(\mathbf{a}^t)$ following (\ref{eq_payoff_inter_game})
 \IF {$o^{t-1}_m(x)=C$}
    \STATE {\bf{if}} {$a^t_{m}\ne\overline{a}^{t-1}_{m}$ and $u^t_{m}(x)\le \overline{u}^{t-1}_{m}(x)$} {\bf{ then }}
    $z^t_{m}(x)\leftarrow z^{t-1}_{m}(x)$ {\bf{end if}}
    \IF {$a^t_{m}\ne\overline{a}^{t-1}_{m}$ and $u^t_{m}(x) > \overline{u}^{t-1}_{m}(x)$}
      \STATE Update the state $z^t_{m}(x)$ with probability
      \begin{equation}
        \label{label_content_trans}
        p_m(z^t_{m}(x)\leftarrow (C, a^t_{m}, u^t_{m}(x)))=\epsilon^{G(u^t_{m}(x)-\overline{u}^{t-1}_{m}(x))}
      \end{equation}
    \ENDIF
    \STATE {\bf{if}} {$a^t_{m}=\overline{a}^{t-1}_{m}$ and $u^t_{m}(x) > \overline{u}^{t-1}_{m}(x)$} {\bf{then}}
       $z^t_{m}(x)\leftarrow (H, \overline{a}^{t-1}_{m}(x), \overline{u}^{t-1}_{m}(x))$ {\bf{end if}}
    \STATE {\bf{if}} {$a^t_{m}=\overline{a}^{t-1}_{m}$ and $u^t_{m}(x) = \overline{u}^{t-1}_{m}(x)$} {\bf{then}}
        $z^t_{m}(x)\leftarrow z^{t-1}_{m}(x)$ {\bf{end if}}
    \STATE {\bf{if}} {$a^t_{m}=\overline{a}^{t-1}_{m}$ and $u^t_{m}(x) < \overline{u}^{t-1}_{m}(x)$} {\bf{then}}
       $z^t_{m}(x)\leftarrow (W, \overline{a}^{t-1}_{m}(x), \overline{u}^{t-1}_{m}(x))$ {\bf{end if}}
 \ELSIF {$o^{t-1}_m(x)=H$}
    \STATE {\bf{if}} {$u^t_{m}(x) > \overline{u}^{t-1}_{m}(x)$} {\bf{then}}
      $z^t_{m}(x)\leftarrow (C, \overline{a}^{t-1}_{m}(x), u^t_{m}(x))$ {\bf{end if}}
    \STATE {\bf{if}} {$u^t_{m}(x) = \overline{u}^{t-1}_{m}(x)$} {\bf{then}}
       $z^t_{m}(x)\leftarrow (C, \overline{a}^{t-1}_{m}(x), \overline{u}^{t-1}_{m}(x))$ {\bf{end if}}
    \STATE {\bf{if}} {$u^t_{m}(x) < \overline{u}^{t-1}_{m}(x)$} {\bf{then}}
        $z^t_{m}(x)\leftarrow (W, \overline{a}^{t-1}_{m}(x), \overline{u}^{t-1}_{m}(x))$ {\bf{end if}}
 \ELSIF {$o^{t-1}_m(x)=W$}
    \STATE {\bf{if}} {$u^t_{m}(x)  > \overline{u}^{t-1}_{m}(x)$} {\bf{then}}
       $z^t_{m}(x)\leftarrow (H, \overline{a}^{t-1}_{m}(x), \overline{u}^{t-1}_{m}(x))$ {\bf{end if}}
    \STATE {\bf{if}} {$u^t_{m}(x)  = \overline{u}^{t-1}_{m}(x)$} {\bf{then}}
      $z^t_{m}(x)\leftarrow (C, \overline{a}^{t-1}_{m}(x), \overline{u}^{t-1}_{m}(x))$ {\bf{end if}}
    \STATE {\bf{if}} {$u^t_{m}(x)  < \overline{u}^{t-1}_{m}(x)$} {\bf{then}}
      \STATE $z^t_{m}(x)\leftarrow (D, \overline{a}^{t-1}_{m}(x), \overline{u}^{t-1}_{m}(x))$ {\bf{end if}}
  \ELSIF {$o^{t-1}_m(x)=D$}
    \STATE Set $z^t_{m}(x)\leftarrow z^{t-1}_{m}(x)$ when ${u}^t_{m}(x)=0$. Otherwise, update $z^t_{m}(x)$ with probability
      \begin{equation}\
        \label{label_discontent_trans}
        \left\{
        \begin{array}{ll}
          p_m(z^t_{m}(x)\leftarrow (C, {a}^{t}_{m}, {u}^t_{m}(x)))=\epsilon^{F(u^t_{m}(x))},\\
          p_m(z^t_{m}(x)\leftarrow (D, \overline{a}^{t-1}_{m}(x), \overline{u}^{t-1}_{m}(x))) =1 - \epsilon^{F(u^t_{m}(x))}
        \end{array}\right.
      \end{equation}
 \ENDIF
 \end{algorithmic}
 \label{alg_trial_and_error}
\end{small}
\end{algorithm}

With the auxiliary states defined in (\ref{label_content_state}), we introduce enhanced trial-and-error learning from \cite{PRADELSKI2012882} in Algorithm~\ref{alg_trial_and_error}. Note that in Algorithm~\ref{alg_trial_and_error}, $G(u)$ and $F(u)$ are strictly monotonically decreasing linear functions for any observed utility $u\in[0,1]$, and the conditions $0<G(u)<1/2$ and $0<F(u)<1/2M$ are to be satisfied (see Theorem~\ref{th_stable_states} for the details).

\section{Analysis of the Regret for Algorithm~\ref{alg_bandit}}
\label{sec_game_analysis}
We note that in Algorithm~\ref{alg_bandit}, the regret is mainly due to sub-optimal actions in the exploration phase and the trial-and-error learning phase. In the latter phase, each player is supposed to learn the optimal matching policies while avoiding collisions in $|\mathcal{X}|$ intermediate games following the rules defined in Algorithm~\ref{alg_trial_and_error}. In each game $\mathcal{G}(x)$ ($x\in\mathcal{X}$), the learning processes of all the players jointly define a large discrete-time Markov chain over the set of all possible auxiliary states (see also~\cite{PRADELSKI2012882}). Therefore, the regret analysis regarding Algorithm~\ref{alg_bandit} is expected to mathematically determine the regret due to the arm-value estimation in the exploration phase and the regret due to the sub-optimal policies derived in Algorithm~\ref{alg_trial_and_error}. For ease of exposition, we first provide the main result of the theoretical bound on the regret of Algorithm~\ref{alg_bandit} in Theorem~\ref{thm_main}, before presenting the analytical procedures for the two phases in concern.
\begin{Theorem}[Main Theorem on the Algorithm Regret]
  Consider a multi-player bandit game with a finite set of contexts, i.e., $|\mathcal{X}|=X$, as defined in Definition~\ref{def_contextual_bandit}. With $T$ rounds of plays and a sufficiently small policy-learning parameter $\epsilon\in[0,1]$ in Algorithm~\ref{alg_bandit}, the regret of Algorithm~\ref{alg_bandit} is upper-bounded by
  \begin{equation}
    \label{eq_bound_regret}
    \Delta R_T \le O(M\log_2^{1+\delta}(T)),
    \end{equation}
    \label{thm_main}
if we set $f(k)=c_1$, $g(k)=c_2k^{\delta}$ ($\delta>1$) and $h(k)=2^k$, where $c_1\ge\frac{16L(L+\eta/3)}{\eta^2}$ with $\eta\in[0,1]$.
\end{Theorem}
\begin{proof}[Proof Sketch]
 Since $r^t_{m,l}\in[0,1]$, the network-wise regret of the exploration phase and the trial-and-error learning phase in the $k$-th epoch can be easily upper-bounded by $M(c_1+c_2k^{\delta})$ in the worst case, in which every player at each round in these two phases produces the maximum regret of 1. Then, to bound the regret in exploitation, we only need to bound the error probability $\Pr\nolimits^k_e$ of the arm-value estimation in the exploration phase and the probability $\Pr\nolimits^k_l$ of learning sub-optimal allocation policies in the trial-and-error learning phase. Thus, we obtain the upper bound of the total regret of all the players in the following form:
 \begin{equation}
   \label{eq_bound_form}
   \Delta R^k\le M(c_1+c_2k^{\delta}) + M(\Pr\nolimits^k_e+\Pr\nolimits^k_l)c_32^k.
 \end{equation}

 The complete proof of Theorem~\ref{thm_main} relies on the analysis of the error probability of the exploration phase in Section~\ref{sub_sec_prob_exploration} and that of the trial-and-error learning phase in Section~\ref{subsec_trial_and_error}. Based on such a two-step analysis, the detail of the proof to Theorem~\ref{thm_main} will be given in Section~\ref{subsec_regret_analysis}.
\end{proof}

\subsection{Error Probability of the Exploration Phase}
\label{sub_sec_prob_exploration}
The goal of the exploration is for every player to obtain the unbiased estimation of the mean values of all arms in each context $x\in\mathcal{X}$. Then, the total sampling period in the exploration phase has to be sufficiently long since the expected sum of regret incurred by the uniformly random exploration of one round for all players can be as large as $O(M)$. Denote $N_m=\vert \mathcal{W}_m\vert$ (cf., Line 6 of Algorithm~\ref{alg_bandit}) as the number of samples accumulated by player $m$ until the end of the current exploration phase in Algorithm~\ref{alg_bandit}.
We note that for a certain policy $\pi_m(x)$ of an individual player $m$, the unbiased estimator of the reward based on the collected reward observation $\mathcal{W}_m$ in Algorithm~\ref{alg_bandit} can be determined using inverse propensity scoring:
\begin{equation}
  \label{eq_ub_estimator}
  \hat{\mu}_m(\pi_m)=\frac{1}{N_m}\sum_{i=1}^{N_m} \frac{\mathbbm{1}(\pi_m(x^i), a^i_{m})v^i_{m}}{1/L},
\end{equation}
where $1/L$ represents the uniformly random action sampling. Let $\hat{\mu}^i_{m}({\pi}_m)=\frac{\mathbbm{1}({\pi}_m(x^i), a^i_{m})v^i_{m}}{1/L}$. Then, we have $E\{\hat{\mu}_m({\pi}_m)\}=E_{(x, r_m)\sim D_m}\{r_{m, {\pi}_m}\}$, and from (\ref{eq_ub_estimator}),
\begin{eqnarray}
  \label{eq_var_bound}
  \begin{array}{ll}
  \displaystyle\mathop{\textrm{Var}}\left\{\hat{\mu}^i_{m}({\pi}_m)\right\} \displaystyle\le E\left\{(\hat{\mu}^i_{m}({\pi}_m))^2\right\}
  =\displaystyle L^2E\left\{{\mathbbm{1}({\pi}_m(x^i), a_{m,i}) (v^i_{m})^2}\right\}
  \displaystyle\le LE\left\{(v^i_{m})^2\right\}\le L.
  \end{array}
\end{eqnarray}
The analysis of the upper bound of the arm-value estimation error relies on two inequalities~\cite{boucheron2013concentration} as follows.
\begin{Fact}[Bernstein Inequality]
  \label{th_Bernstein}
  If for a sequence of random variables $Y_1,\ldots, Y_N$, $\Pr(|Y_i| \le c) = 1$ and $E(Y_i) = 0$, then for any $N > 0$,
  \begin{equation}
    \label{eq_Bernstein}
    \Pr(\frac{1}{N}\sum_{i=1}^{N}Y_i\ge\eta)\le 2 \exp(-\frac{N\eta^2}{2\sigma^2+2c\eta/3}),
  \end{equation}
  where $c$ is a constant and $\sigma^2=\frac{1}{N}\sum_{i=1}^{N}\mathop{\textrm{Var}}({Y_i})$.
\end{Fact}
\begin{Fact}[Chernoff Inequality]
  \label{th_Chernoff}
  If for a sequence of random variables $Y_1,\ldots, Y_N$, $|Y_i| \le 1$, then for any $N > 0$ and $0<\eta<1$,
  \begin{equation}
    \label{eq_Chernoff}
    \Pr(\sum_{i=1}^{N}Y_i\!\le \!(1\!-\!\eta) E\left\{\sum_{i=1}^{N}Y_i\right\})\!\le\! \exp(-\eta^2E\left\{\sum_{i=1}^{N}Y_i\right\}\!/2).
  \end{equation}
\end{Fact}
Based on (\ref{eq_var_bound})-(\ref{eq_Chernoff}) we obtain Lemma~\ref{la_lamma_exploreation_error} as follows.
\begin{Lemma}
\label{la_lamma_exploreation_error}
With Algorithm~\ref{alg_bandit}, all the players have a sufficiently accurate arm-value estimation after $T_0$ explorations, with $T_0$ given by:
\begin{equation}
  \label{eq_Lemma_1_T0}
  T_0\ge \max\left(16LX\frac{L+c\eta/3}{\eta^2}\ln\left(\frac{4ML}{\gamma}\right), 32L\ln\left(\frac{2M}{\gamma}\right)\right).
\end{equation}
where $\gamma$ is the pre-determined exploration error probability for a maximum estimation error $\eta$.
\end{Lemma}
\begin{proof}

For player $m$ which has undergone at least $C$ rounds of valid explorations (i.e., explorations with no collisions), the probability of not having sufficiently accurate arm-value estimations for a non-colliding policy $\pi_m$ ($\forall m\in\mathcal{M}$) adopted in the exploitation phase in Algorithm~\ref{alg_bandit} is bounded by
  \begin{equation}
    \label{eq_estimation_bound}
    \begin{array}{ll}
      \Pr\left( \sup\limits_{m\in{\mathcal{M}}, \pi_m}\left\{\left(\hat{\mu}_{m}(\pi_m)\!-\!E\{r_{m,\pi_m}\}\right)\!>\!\eta \Big\vert \forall m: \vert W_m\vert \!\ge\! C \right\}\right)\\
      \stackrel{\textrm{(a)}}{\le}\displaystyle\sum\limits_{m=1}^{M}\sum\limits_{\pi_m\in{\Pi_m}}\Pr\left(\hat{\mu}_{m}(\pi_m)-E\{r_{m,\pi_m}\}>\eta \Big\vert \vert W_m\vert \ge C\right)\\
      \stackrel{\textrm{(b)}}{\le}\displaystyle\sum\limits_{m\!=\!1}^{M}\! \sum\limits_{\pi_m\in{\Pi_m}}\!\!
      \frac{\sum\limits_{N_m\!=\!C}^{\infty}\!\Pr\left(\hat{\mu}_{m}(\pi_m)\!-\!E\{r_{m,\pi_m}\}\!>\!\eta
       \big\vert \vert W_m\vert \!=\! N_m \right)\!\Pr(\vert W_m\vert \!=\! N_m)}{\Pr\left(\vert W_m\vert\ge C\right)} \\
       {\le}\displaystyle\sum\limits_{m\!=\!1}^{M}\sum\limits_{\pi_m\in{\Pi_m}}\sum\limits_{N_m\!=\!C}^{\infty}
      \Pr\left(\hat{\mu}_{m}(\pi_m)\!-\!E\{r_{m,\pi_m}\}\!>\!\eta \Big\vert \vert W_m\vert\!=\!N_m \right)
         \times\Pr(\vert W_m\vert \!=\! N_m \Big\vert \vert W_m\vert\ge C) \\
      \stackrel{\textrm{(c)}}{\le}\displaystyle\sum\limits_{m\!=\!1}^{M}\sum\limits_{\pi_m\in{\Pi_m}}\!\sum\limits_{N_m\!=\!C}^{\infty} e^{-\frac{N_m\eta^2}{2L+2c\eta/3}} \Pr(\vert W_m\vert \!=\! N_m \Big\vert \vert W_m\vert\ge C)\\
      {\le}2ML^X e^{-\frac{C\eta^2}{2L+2c\eta/3}}\displaystyle\sum\limits_{N=C}^{\infty} \Pr(\vert \mathcal{W}\vert\!=\!N_m\Big\vert N_m\!\ge\! C)
      \le 2ML^X\exp\left(-\displaystyle\frac{C\eta^2}{2L+2c\eta/3}\right),
    \end{array}
  \end{equation}
  where $\Pi_m$ is the set of deterministic policies for player $m$ and $|\Pi_m|=L^X$, (a) is obtained by the union bound, (b) is obtained following the Partition Theorem and (c) is obtained following the Bernstein Inequality in Fact~\ref{th_Bernstein}. To satisfy the condition of sufficient accuracy $\eta$ with an error probability $\gamma_1$, we have
  \begin{equation}
    \label{eq_estimation_bound_on_rounds}
    \begin{array}{ll}
      2ML^X\exp\left(-\displaystyle\frac{C\eta^2}{2L+2c\eta/3}\right)\le \gamma_1
       \Rightarrow C\ge\displaystyle\frac{2L+2c\eta/3}{\eta^2}\ln\left(\frac{2ML^X}{\gamma_1}\right).
    \end{array}
  \end{equation}

  Note that the above condition in (\ref{eq_estimation_bound_on_rounds}) is obtained when the players sample the arms uniformly at random and no collision occurs. To obtain the condition for accumulating sufficiently large number of valid arm observations for each player, we denote $A^i_{m}$ as the event that a player $m$ observes any arm $l\in\mathcal{A}_m$ without experiencing collision at the $i$-th sample. During the exploration phase, whether experiencing a collision is independent of the context that the game is in. Then, $\forall i=1,2,\ldots$, we have $\Pr(A^i_{m})=(1-\frac{1}{L})^{M-1}$. For a sequence of $N$ i.i.d. samples $\{A_m^i\}_{i=1}^N$, we have
  \begin{equation}
    \label{eq_bound_on_total_rounds}
    \begin{array}{ll}
      \Pr(\exists m \textrm{ s.t. }\sum\limits_{i=1}^N A^i_{m}\le \displaystyle\frac{N}{2}E\{A^i_{m}\})
      \stackrel{\textrm{union bound}}{\le} \sum\limits_{m=1}^M\Pr(\sum\limits_{i=1}^N A^i_{m}\le \displaystyle\frac{N}{2}E\{A^i_{m}\})
      \stackrel{\textrm{Fact 2}}{\le}M\exp\left(-\displaystyle\frac{N}{8}(1-\frac{1}{L})^{M-1}\right).
    \end{array}
  \end{equation}
  For the probability in (\ref{eq_bound_on_total_rounds}) to be upper-bounded by $\gamma_2$, we need
  \begin{equation}
    \label{eq_adjusted_bound_on_total_rounds}
    \begin{array}{ll}
      M\exp\left(-\displaystyle\frac{N}{8}(1-\frac{1}{L})^{M-1}\right)\le \gamma_2
      \Rightarrow \displaystyle N\ge8(1-\frac{1}{L})^{-(M-1)}\ln\left(\frac{M}{\gamma_2}\right).
    \end{array}
  \end{equation}
  Then, with probability $(1-\gamma_2)$, we have $\forall m\in\mathcal{M}$, $\sum_{i=1}^N A^i_{m}\ge\frac{N}{2}E\{A^i_{m}\}$. To ensure that every player has a sufficient number of valid observations, we also need
 \begin{equation}
   \label{eq_final_bound_on_total_exploretaion_rounds}
   \begin{array}{ll}
     \displaystyle\frac{N}{2}E\{A^i_{m}\}
     \ge C \stackrel{\textrm{(\ref{eq_estimation_bound_on_rounds})}}{\ge} \frac{2L+2c\eta/3}{\eta^2}\ln\left(\frac{2ML^X}{\gamma_1}\right)
     \Rightarrow N\ge 2 (1-\frac{1}{L})^{-(M-1)}\displaystyle\frac{2L+2c\eta/3}{\eta^2}\ln\left(\frac{2ML^X}{\gamma_1}\right).
   \end{array}
 \end{equation}
 Since for any $L>1$, $(1-\frac{1}{L})^{M-1}\ge\frac{1}{4L}$, we have
 \begin{equation}
   \label{eq_Lemma_1}
   N\ge \max\left(16L\frac{L+c\eta/3}{\eta^2}\ln\left(\frac{2ML^X}{\gamma_1}\right), 32L\ln\left(\frac{M}{\gamma_2}\right)\right).
 \end{equation}

 Let the event $\sup_{m, \pi_m}\left(\hat{\mu}_{m}(\pi_m)\!-\!E\{r_{m,\pi_m}\}\right)\!>\!\eta$ be denoted by $A$, and the event $\forall m: \vert \mathcal{W}_m\vert\ge C$ be denoted by $B$. Then,  (\ref{eq_estimation_bound}) provides the upper bound of $\Pr(A|B)$, and (\ref{eq_bound_on_total_rounds}) leads to the upper bound of ${\Pr}(\overline{B})$. To guarantee that all players have satisfactory estimation errors of $\eta$ for each arm, we have the following bound:
 \begin{equation}
   \label{eq_error_deduction}
  \begin{array}{ll}
    \Pr({A}) = \Pr(A|B)\Pr(B) + \Pr(A|\overline{B})\Pr(\overline{B})
    \le \Pr({A}|B) + \Pr(\overline{B}) = \gamma_1+\gamma_2.
  \end{array}
 \end{equation}
 Then, having $\gamma_1=\gamma_2=\gamma/2$, (\ref{eq_Lemma_1}) guarantees that with more than $N$ rounds of exploration, any policy is estimated with an error within $\eta$ with probability $1-\gamma$. This leads to (\ref{eq_Lemma_1_T0}).
\end{proof}

From (\ref{eq_Lemma_1_T0}) in Lemma~\ref{la_lamma_exploreation_error}, we note that for an error probability of arm-value estimation with maximum bias $\eta$, $\Pr^k_e=\gamma$, Algorithm~\ref{alg_bandit} needs to undergo at least $T_0$ rounds of exploration as
\begin{equation}
  \label{eq_min_T_for_exploration}
  T_0 = 16L\frac{L+c\eta/3}{\eta^2}\ln\left(\frac{4ML^X}{\gamma}\right).
\end{equation}
If the exploration has at least $c_1=\frac{16L(L+c\eta/3)}{\eta^2}$ turns at each epoch, then, at $k$-th epoch, for a maximum estimation error $\eta$ the error probability can be bounded as follows
\begin{equation}
  \label{eq_bound_of_exploration_probability}
  \begin{array}{ll}
    c_1k = \displaystyle\frac{16L(L+c\eta/3)}{\eta^2}k \ge \frac{16L(L+c\eta/3)}{\eta^2}\ln\left(\frac{4ML^X}{\gamma}\right)
    \Rightarrow \displaystyle\Pr\nolimits^k_e=\gamma\le 4ML^Xe^{-k}.
  \end{array}
\end{equation}
Note that with the normalized arm-values, we can simply choose $c=1$ in (\ref{eq_bound_of_exploration_probability}).

\subsection{Error Probability in the Trial-and-error Phase}
\label{subsec_trial_and_error}
In addition to Lemma~\ref{la_lamma_exploreation_error} and (\ref{eq_bound_of_exploration_probability}), we need to further analyze the impact of the arm-value estimation errors on the learning results in the trial-and-error phase. Specifically, we expect that the optimal contextual bipartite matching policy derived based on the biased arm-value estimation is the same as the optimal policy derived based on the real expected arm-values. Lemma~\ref{la_good_allocation} confirms this presumption.
\begin{Lemma}
  \label{la_good_allocation}
  Assume that the expected reward estimated by player $m$ for a non-colliding policy $\pi_m$, $\hat{\mu}_{m, \pi_m}(x)$, satisfies $\vert \hat{\mu}_{m, \pi_m}(x) - E\{r_{m,\pi_m}|x\} \vert\le\eta$. Consider two intermediate games (cf., Definition~\ref{def_inter_game}): $\mathcal{G}(x)$, which is constructed upon the real expected arm-values $E\{r_{m,l}|x\}$, and $\hat{\mathcal{G}}(x)$, which is constructed upon the estimated arm-values $\hat{\mu}_{m,l}$, respectively. For $\mathcal{G}(x)$ where $V_{\pi}(x)=\sum_{m=1}^{M}E\{r_{m,\pi_m}|x\}$, we denote an optimal joint policy as $\pi^*$ and a best non-optimal joint policy as $\tilde{\pi}$. Then, if
  \begin{equation}
    \label{eq_discernable_condition}
    \eta<\frac{{V}_{\pi^{*}}(x)-{V}_{\tilde{\pi}}(x)}{2M},
  \end{equation}
  for $\hat{\mathcal{G}}(x)$ where $\hat{V}_{\pi}(x)=\sum_{m=1}^{M}\hat{\mu}_{m, \pi_m}(x)$, we have $\hat{V}_{\pi^*}(x)\!>\!\hat{V}_{\tilde{\pi}}(x)$ as well.
\end{Lemma}
\begin{proof}
  Since $L\ge M$, $\pi^*$ and $\tilde{\pi}$ must be collision-free. By inequality construction of the condition at the beginning of Lemma~\ref{la_good_allocation} and the definitions of $V_{\pi}(x)$ and $\hat{V}_{\pi}(x)$, we have (note that we omit $x$ for conciseness)
  \begin{equation}
    \label{eq_sum_bias_bound}
    -M\eta\le\sum_{m=1}^{M}\left(\hat{\mu}_{m,\pi_m}-E\{r_{m,\pi_m}\}\right) \le M\eta.
  \end{equation}
  Then, for $\pi^*$ we have
  \begin{equation}
    \label{eq_biased_optimal_value}
    \sum_{m=1}^{M}\hat{\mu}_{m,\pi^*_m}  \ge \sum_{m=1}^{M}E\{r_{m,\pi^*_m}\} - M\eta = {V}_{\pi^*} - M\eta.
  \end{equation}
  For any non-optimal policy $\pi$, its value can be bounded by the best non-optimal policy $\tilde{\pi}$ as follows:
  \begin{equation}
    \label{eq_biased_second_optimal_value}
     \sum_{m=1}^{M}\hat{\mu}_{m,\pi_m}  \!\le\! \sum_{m=1}^{M}E\{r_{m,\pi_m}\} +\!M\eta \!=\! {V}_{\pi} \!+\! M\eta \!\le\! {V}_{\tilde{\pi}}+M\eta.
  \end{equation}
  Subtracting (\ref{eq_biased_optimal_value}) by (\ref{eq_biased_second_optimal_value}), we obtain the following inequality with the condition $\eta<\frac{{V}_{\pi^{*}}(x)-{V}_{\tilde{\pi}}(x)}{2M}$:
  \begin{equation}
    \label{eq_condition_biased_optimal_value}
    \hat{V}_{\pi^*} - \hat{V}_{\tilde{\pi}} \ge {V}_{\pi^*} - {V}_{\tilde{\pi}}-2M\eta>0.
  \end{equation}
  (\ref{eq_condition_biased_optimal_value}) shows that for game $\hat{\mathcal{G}}(x)$, any optimal joint policy $\pi^{*}(x)$ in game $\mathcal{G}(x)$ also achieves strictly higher social reward than the non-optimal policies in $\mathcal{G}(x)$. Therefore, a social optimal policy in game $\hat{\mathcal{G}}(x)$ must also be optimal in $\mathcal{G}(x)$.
\end{proof}

Lemmas~\ref{la_lamma_exploreation_error} and~\ref{la_good_allocation} guarantee that as long as the estimation error $\eta$ is small enough, the true social optimal policy can be derived based on the biased estimation of arm-values after sufficiently long exploration in Algorithm~\ref{alg_bandit}. Therefore, we only need to examine the policy efficiency of the trial-and-error learning phase based on the rules defined in Algorithm~\ref{alg_trial_and_error}. Regarding the intermediate game $\mathcal{G}(x)$ in context $x$, we have
\begin{Lemma}\label{lemma_existence}
    The social-optimal payoff by the players in game $\mathcal{G}(x)$ is achieved at a pure NE.
\end{Lemma}
\begin{proof}
  Lemma~\ref{lemma_existence} relies on the assumption of $L\ge M$. We note that at epoch $k$, game $\mathcal{G}(x)$ with fixed arm values for $m\in\mathcal{M}$, $\mu^k_{m,l}(x)$, belongs to the category of one-sided matching games with user preferences~\cite{hylland1979efficient}. Then, by randomly ordering the players in a list, and sequentially assigning each player in the list their best available arm, we are able to obtain a non-colliding allocation $\mathbf{a}^k=[a^k_1,\ldots, a^k_M]^{\top}$. It is straightforward to check that $\forall m\in\mathcal{M}$, $a^k_m$ is a best response to the joint actions of the other players $a^k_{-m}$.Thus, $\mathbf{a}^k$ constructs a pure NE, and we know that more than one pure NE exists in $\mathcal{G}(x)$.

  Furthermore, with $L\ge M$, player $m$'s better response to $a^k_{-m}$ can only be pulling a free arm. Indeed, a player's better response always leads to a Pareto improvement, since no other players changes their payoffs. Then, we can check by contradiction that the social optimal policy $\mathbf{a}^{k,*}$ in $\mathcal{G}(x)$, where $V^x(\mathbf{a}^{k,*})=\max\limits_{\mathbf{a}}\sum_{m=1}^{M}u^{x,k}_m(\mathbf{a})$, is also an NE. Firstly, with $L\ge M$, the optimal action $\mathbf{a}^{k,*}$ has no collision. Otherwise, a colliding player can always find a free arm as the better response, which constitutes a Pareto improvement. Secondly, at $\mathbf{a}^{k,*}$ no player is able to find a better response. Otherwise, at least one player $m$ can find some free arm $a'_m$, that leads to a joint action $\mathbf{a}'=({a}'_m, {a}^{k,*}_{-m})$ s.t. $V^x(\mathbf{a}^{k,*})<V^x(\mathbf{a}')=\sum_{i\ne m}u^{x,k}_i({a}^{k,*}_i)+u^{x,k}_m({a}'_m)$, contradicting with the optimality assumption. Therefore, by the definition of an NE, we obtain Lemma~\ref{lemma_existence}.
\end{proof}

With Lemma~\ref{lemma_existence}, we are left to show that the policies obtained from Algorithm~\ref{alg_trial_and_error} converge to not only an NE, but also the most efficient NE of the intermediate game. Note that following the rules of state transition defined in Algorithm~\ref{alg_trial_and_error}, the state-updating dynamics of each player $m$ jointly constitute a large discrete-time Markov chain over the set of the joint auxiliary states $\mathbf{z}(x)=[z_1(x),\ldots,z_M(x)]^{\top}$ as defined in (\ref{label_content_state}). Following the approach of the Markov chain-based analysis for log-linear learning in~\cite{PRADELSKI2012882}, we are able to examine the efficiency of the trial-and-error learning phase in Algorithm~\ref{alg_bandit} for a given intermediate game $\mathcal{G}(x)$. Before proceeding, we introduce the concepts of regular perturbation and stochastically stable states from~\cite{PRADELSKI2012882,10.2307/2951778} for Markov chains in Definitions~\ref{def_reg_perturbation} and~\ref{def_stochastic_stability}.
\begin{Definition}[Regular Perturbation]
  \label{def_reg_perturbation}
  Let $P^0$ denote the transition matrix for a stationary Markov chain over a finite state space $\mathcal{Z}$, and $P^{\epsilon}$ ($\epsilon\ne0$) be a family of perturbed Markov chains corresponding to $P^0$, where $\epsilon$ is a scalar measuring the perturbation level. $P^{\epsilon}$ is a regular perturbation of $P^0$ if (a) $P^{\epsilon}$ is ergodic for all sufficiently small $\epsilon$, (b) $\lim\limits_{\epsilon\rightarrow\infty}P^{\epsilon}(\mathbf{z},\mathbf{z}')=P^{0}(\mathbf{z},\mathbf{z}')$, and (c) $P^{\epsilon}(\mathbf{z},\mathbf{z}')>0$ for $\epsilon$ implying that $\exists r(\mathbf{z},\mathbf{z}')$ s.t. $0< \lim\limits_{\epsilon\rightarrow0} P^{\epsilon}(\mathbf{z},\mathbf{z}')/\epsilon^{-r(\mathbf{z},\mathbf{z}')}<\infty$. The function $r(\mathbf{z},\mathbf{z}')$ is known as the resistance of transition $\mathbf{z}\rightarrow\mathbf{z}'$.
\end{Definition}

\begin{Definition}[Stochastically Stable States]
  \label{def_stochastic_stability}
  Let $P^{\epsilon}$ be a regular perturbation of $P^0$ and $p_{\epsilon}$ be its unique stationary distribution. $\mathbf{z}\in\mathcal{Z}$ is a stochastically stable state iff $\lim\limits_{\epsilon\rightarrow0}p_{\epsilon}(\mathbf{z})>0$.
\end{Definition}

Recall the action updating rule for a content player in (\ref{label_cotent_prob}) and the auxiliary state updating rules for a content/discontent player in (\ref{label_content_trans}) and (\ref{label_discontent_trans}) of Algorithm~\ref{alg_trial_and_error}. We obtain the unperturbed Markov process $P^0$ for the joint auxiliary states $\mathbf{z}(x)$ in game $\mathcal{G}(x)$ by setting $\epsilon=0$ in (\ref{label_cotent_prob}), (\ref{label_content_trans}) and (\ref{label_discontent_trans}). Alternatively, for a very small $\epsilon$ ($\epsilon>0$), we obtain the perturbed Markov chain $P^{\epsilon}$, where the larger the exponents of $\epsilon$  is in (\ref{label_content_trans}) and (\ref{label_discontent_trans}) (e.g., $G(u_m^t(x)-\overline{u}_m^{t-1}(x))$ and $F(u_m^t(x))$), the smaller the potential transition probability of $P^{\epsilon}$ is from a current joint state to another reachable state in Algorithm~\ref{alg_trial_and_error}. Therefore, we know that the resistance of a feasible transition (cf., Definition~\ref{def_reg_perturbation}) is partially determined by the values of $G(u_m^t(x)-\overline{u}_m^{t-1}(x))$ and $F(u_m^t(x))$ of each link $m$. This paves the way for identifying the stochastically stable states of the perturbed Markov chain $P^{\epsilon}$ by analyzing the rooted trees spanned from the directed graph with the vertices and edges corresponding to the joint states and feasible transitions of the Markov chain $P^{\epsilon}$.
By~\cite{PRADELSKI2012882}, we know that for a perturbed Markov process $P^{\epsilon}$ with a set of stochastically stable states $\mathcal{Z}^*$, there exists $\epsilon_{\alpha} > 0$ for any small $\alpha > 0$ s.t. whenever $0 < \epsilon\le  \epsilon_\alpha$, $\mathbf{z}(t)\in\mathcal{Z}^*$ for at least $1-\alpha$ of all periods in the process. Therefore, it is natural to desire that the social optimal NE of a game $\mathcal{G}(x)$ (see Lemma~\ref{lemma_existence}) constitute the stochastically stable states of the Markov process defined by the rules given in Algorithm~\ref{alg_trial_and_error} when $\epsilon>0$. This is guaranteed by the following theorem.
\begin{Theorem}
  \label{th_stable_states}
  Suppose that all players in an intermediate game $\mathcal{G}(x)$ follow the updating rules in Algorithm~\ref{alg_trial_and_error} with the experimentation parameter $\epsilon$ and the acceptance functions $F(u)$ and $G(u)$ using the same parameters $\alpha_{11},\alpha_{21}>0$, $\alpha_{12}$ and $\alpha_{22}$, s.t. $0<G(u)<1/2$ and $0<F(u)<1/2M$:
  \begin{equation}
    \label{eq_accpt_functions}
    \left\{
    \begin{array}{ll}
      F(u)=-\alpha_{11}u+\alpha_{12}\\
      G(u)=-\alpha_{21}u+\alpha_{22}.
    \end{array}\right.
  \end{equation}
  Then, based on (\ref{label_content_state}), every stochastically stable state $\mathbf{z}^*(x)=(z^*_1(x),\ldots, z^*_M(x))$ maximizes the social welfare of the game in the form of $\sum_{m=1}^M u^x_m(\overline{a}^*_m(x))$, where $\overline{a}^*_m(x)$ is the benchmark action in $z^*_m(x)$ and constitutes an NE in $\mathcal{G}(x)$.
\end{Theorem}

\begin{proof}
  Lemma~\ref{lemma_existence} guarantees that the social optimal policy of game $\mathcal{G}(x)$ is also a pure NE. Therefore, the learning scheme defined in Algorithm~\ref{alg_trial_and_error} satisfies condition (i) of~\cite[Theorem 1]{PRADELSKI2012882}. Let $P^{\epsilon}(x)$ denote the family of (perturbed) Markov processes defined in Algorithm~\ref{alg_trial_and_error} in a single epoch for context $x$. Following the same approach of proving~\cite[Theorem 1]{PRADELSKI2012882}, we only need to show that the social optimal NE are stochastically stable states of $P^{\epsilon}(x)$, namely
  \begin{itemize}
    \item [(a)] these social welfare-maximizing NE are aligned with some states contained in the recurrent communication classes of the unperturbed process $P^0(x)$, and
    \item [(b)] in the sub-graph of states constructed over the directed transitions between the recurrence classes of $P^0(x)$, these NE minimize the stochastic potential (see~\cite{10.2307/2951778} for the formal definition). Namely, there exists a state-tree spanned on each NE state that minimizes the sum resistance of the edges (see also Definition~\ref{def_reg_perturbation}) in the tree among all the possible spanning trees in this recurrence graph.
  \end{itemize}
  Condition (a) relies on the identification of recurrence classes (cf.~\cite[Lemma 1]{PRADELSKI2012882}). Condition (b) requires enumerating the minimum  resistance of the edges ended on different states in the considered sub-graph (cf.~\cite[Lemmas 2-6]{PRADELSKI2012882}).

  The proof to Theorem~\ref{th_stable_states} strictly follows the approach of proof to~\cite[Theorem 1]{PRADELSKI2012882}, except the slight difference in the interdependence property\footnote{By~\cite{PRADELSKI2012882}, a multi-player game is interdependent if for any joint action $\mathbf{a}$ and any subset of players $\mathcal{J}$, there exists some player $i\notin \mathcal{J}$ and two joint actions of $\mathcal{J}$, $\mathbf{a}_{\mathcal{J}}$ and $\mathbf{a}'_{\mathcal{J}}$, s.t. when fixing the actions of the players not in $\mathcal{J}$, player $i$'s payoff w.r.t. $\mathbf{a}_{\mathcal{J}}$ and $\mathbf{a}'_{\mathcal{J}}$ are different.} between game $\mathcal{G}(x)$ and the non-cooperative game considered in~\cite{PRADELSKI2012882}. In $\mathcal{G}(x)$, we note that a player $i$ can only cause another non-colliding, non-experimenting player $j$'s payoff to decrease or remain the same (both with non-zero probability) by altering its own action. Such a ``partial interdependence'' property indicates that only the sub-set of non-colliding players are interdependent on the action of the other players due to potential collision. This eliminates any path in a state graph of the Markov process $P^{\epsilon}(x)$ s.t. a non-colliding, non-experimenting player $j$'s state transits to mood $o_j=H$ (i.e., observing reward increase) due to another player $i$'s action experimentation. Therefore, we only need to consider the $O(1)$ probability that one player's experimenting action collides with another player in the original proof to~\cite[Lemma 1]{PRADELSKI2012882} and obtain the following result:
  \begin{Proposition}[Lemma 1 in~\cite{PRADELSKI2012882}]
    \label{lemma_recurrent_class}
    Denote by $\mathcal{D}_0(x)$ the set of the joint states $\mathbf{z}(x)\in\mathcal{Z}(x)$, where $\forall m, o_m(x)=D$ and $\mathcal{C}_0(x)$ the set of $\mathbf{z}(x)$ where $\forall m, o_m(x)=C$. The recurrence classes of the unperturbed Markov process $P^0(x)$ are $\mathcal{D}_0(x)$ (as a single state) and every singleton $\mathbf{z}(x)\in\mathcal{C}_0(x)$.
  \end{Proposition}

  Following the approach of the proof to~\cite[Theorem 1]{PRADELSKI2012882}, we denote $\mathcal{E}_0(x)$ as the subset of $\mathcal{C}_0(x)$ where the benchmark actions align with a pure NE. Then, to analyze the minimum resistance of an edge out-going from $\mathbf{z}_E(x)\in\mathcal{E}_0(x)$ in the transition graph of the recurrence states, we only need to consider a single case regarding the path between $\mathbf{z}_E(x)$ and $\mathbf{z}_D(x)\in\mathcal{D}_0(x)$. Due to partial interdependence in $\mathcal{G}(x)$, one single player experimenting two consecutive times can only lead to a path of transitions $C\rightarrow W\rightarrow D$ due to twice collisions with a probability of $O(\epsilon^2)$. Since any state $\mathbf{z}(x)$ with at least one player being discontent has 0 resistance to $\mathcal{D}_0(x)$~\cite[Claim 1]{PRADELSKI2012882}, this leads to a simplified version of the proof to~\cite[Lemma 2]{PRADELSKI2012882} and thus the following proposition:
  \begin{Proposition}[Lemma 2 in \cite{PRADELSKI2012882}]
    \label{prop_lemma_2}
    In the state graph of perturbed transitions constructed on the recurrence classes of $P^{0}(x)$, $\forall \mathbf{z}_e(x)\in\mathcal{E}_0(x)$, $\mathbf{z}_e(x)\rightarrow \mathcal{D}(x)$ is an easy edge. Namely, $\mathbf{z}_e(x)\rightarrow \mathcal{D}(x)$ has a minimized resistance of 2 among all possible out-going edges from $\mathbf{z}_e(x)$.
  \end{Proposition}

  The rest part of the proof follows exactly the same approach of the proof to~\cite[Theorem 1]{PRADELSKI2012882}, where the resistance of edges out-going from both non-equilibrium content states and discontent states are also identified, and then the easy trees (i.e., those with the minimum sum of resistance) are constructed on each recurrence state. Since we do not need to make any change to the intermediate proofs to~\cite[Lemmas~3-6]{PRADELSKI2012882}, for conciseness, we omit the details of the proof and suggest the readers to refer to~\cite[Section 6]{PRADELSKI2012882}. Because~\cite[Theorem 1]{PRADELSKI2012882} holds, by Lemma~\ref{lemma_existence} we know that the stochastically stable states of $P^{\epsilon}$ coincide with the social optimal NE strategies of the considered game, which completes the proof to Theorem~\ref{th_stable_states}.
\end{proof}

Together with Lemma~\ref{la_good_allocation}, Theorem~\ref{th_stable_states} indicates that for the intermediate game $\mathcal{G}(x)$ constructed directly upon the estimated arm-values $\mu^k_{m,l}$, we can always find an $\epsilon_{\alpha}$ and a sufficiently large number of rounds s.t. each player visits the real social optimal actions of the underlying bandit game for at least $1-\alpha$ of the total trial-and-error rounds. However, if $\mathcal{G}(x)$ has multiple social optimal NE\footnote{We can construct such a game by setting the expected rewards of $M$ arms to be uniformly $0<\overline{\mu}<1$ for each player and the other arms to be always 0, with the non-zero arm-values sampled from discrete distribution.}, the non-cooperative players may reach a sub-optimal joint allocation with solely the action selection scheme in Line 24 of Algorithm~\ref{alg_bandit}. We overcome this uncertainty by replacing the estimated arm-values in $\mathcal{G}(x)$ with the randomly perturbed values $\tilde{\mu}^k_{m,l}(x)={\mu}^k_{m,l}(x) + \xi_{m,l}(x)$, where $\xi_{m,l}(x)$ is independently sampled following a uniform distribution over $[-\xi, \xi]$ for context $x$.

Therefore, we obtain a condition that $\forall l\in\mathcal{A}_m, \vert {\mu}^k_{m,l}(x) - \tilde{\mu}^k_{m,l}(x) \vert\le\xi$. Applying the same approach of proving Lemma~\ref{la_good_allocation}, we can always find a sufficiently small $\xi$, s.t. for the optimal policy $\pi^*$ and the best non-optimal policy $\tilde{\pi}$, the following inequality is satisfied
\begin{equation}
  \label{eq_discernable_codition2}
  \xi<\frac{\sum\limits_{m=1}^M\left(\mu^k_{m,\pi^*_m}(x)-\mu^k_{m,\tilde{\pi}_m}(x)\right)}{2M}.
\end{equation}
Thereby, any optimal NE policy of game $\mathcal{G}(x)$ becomes the candidate optimal NE policies of the new game $\tilde{\mathcal{G}}(x)$ constructed upon the perturbed arm-value $\tilde{\mu}^k_{m,l}(x)$. We consider two different and non-colliding actions $\mathbf{a}$ and $\mathbf{a}'$ s.t. they achieve equal social rewards in $\mathcal{G}(x)$, i.e., $\sum_{m\in\mathcal{M}}{\mu}^k_{m,a_m}=\sum_{m\in\mathcal{M}}{\mu}^k_{m,a'_m}$. Omitting $x$ again, we consider the probability that $\mathbf{a}$ and $\mathbf{a}'$ also achieve the same social reward in game $\tilde{\mathcal{G}}(x)$:
\begin{equation}
  \label{eq_zero_prob_equal_payoff}
  \begin{array}{ll}
    \Pr(\sum\limits_{m\in\mathcal{M}}\tilde{\mu}^k_{m,a_m}=\sum\limits_{m\in\mathcal{M}}\tilde{\mu}^k_{m,a'_m})\\
  =\Pr(\sum\limits_{m\in\mathcal{M}}\left(\hat{\mu}^k_{m,a_m}+\xi_{m,a_m}\right)=\sum\limits_{m\in\mathcal{M}}\left(\hat{\mu}^k_{m,a'_m}+\xi_{m,a'_m}\right))
    =\Pr(\sum\limits_{m\in\mathcal{M}}(\xi_{m,a_m}-\xi_{m,a'_m})=0).
  \end{array}
\end{equation}
Since at least one player $i\in\mathcal{M}$ adopts different actions $a_i\ne a'_i$, $\sum\limits_{m\in\mathcal{M}}(\xi_{m,a_m}-\xi_{m,a'_m})$ is a sum of at least two independent continuous random variables. Then, we have $\Pr(\sum\limits_{m\in\mathcal{M}}(\xi_{m,a_m}-\xi_{m,a'_m})=0)=0$. Therefore, the perturbation $\forall m,l: \xi_{m,l}(x)$ guarantees that the social optimal SE of $\tilde{\mathcal{G}}(x)$ is unique with probability 1. This leads to the operation in Line 11 of Algorithm~\ref{alg_bandit}\footnote{Since the gaps between the optimal and the secondary social optimal rewards is not known in advance, we adopt a decaying factor ${1}/{k}$ for the perturbation in Algorithm~\ref{alg_bandit}.} and Proposition~\ref{prop_unique_NE}.
\begin{Proposition}
  \label{prop_unique_NE}
  With a sufficiently small perturbation parameter $\xi$ satisfying (\ref{eq_discernable_codition2}), the trial-and-error phase in Algorithm~\ref{alg_bandit} reaches a unique social-optimal NE of the intermediate game with probability 1.
\end{Proposition}

We consider $g(k)=T_1$ rounds of plays in the $k$-th trial-and-error learning phase, which contains $X$ independent perturbed Markov processes. Now, we are ready to examine the inherent error probability of not reaching stochastically stable states in $P^{\epsilon}(x)$. Suppose that each process $P^{\epsilon}(x)$ continues for $T_1(x)$ rounds, then we have $\sum_{x\in\mathcal{X}}T_1(x)=T_1$. We denote $\mathcal{E}^*(x)$ the singleton of stochastically stable state that aligns with the unique social optimal NE in context $x$. Then, with Line 24 of Algorithm~\ref{alg_bandit}, the probability of players selecting optimal actions in the exploitation phase is determined by the frequency that $\forall x\in\mathcal{X}:\mathcal{E}^*(x)$ are visited. We denote $A_x$ as the event that for context $x$ the optimal policy is adopted after the trial-and-error phase and $A$ the event that for all contexts the optimal policies are adopted. Then, we have
\begin{equation}
  \label{eq_probability_bound_good_actions}
  \begin{array}{ll}
  \Pr(A) = 1 -\Pr(\overline{A})
  \ge 1 -\Pr(\mathop{\cup}\limits_{x\in\mathcal{X}}\overline{A}_x)\stackrel{\textrm{union bound}}{\ge}1-\sum\limits_{x\in\mathcal{X}}\Pr(\overline{A}_x).
\end{array}
\end{equation}

To bound $\Pr(\overline{A}_x)$, we apply the approach of analyzing the accumulated weights of random walks on general (irreversible) finite-state Markov chains from~\cite{chung_et_al_2912}. At epoch $k$, with the initialization step in Line 12 of Algorithm~\ref{alg_bandit}, the trial-and-error learning process for a game $\mathcal{G}(x)$ constitutes a random walk of $T_1(x)$ steps with an arbitrary initial distribution $\phi(x)$ over the states on the Markov process $P^{\epsilon}(x)$. Let $\mathbbm{1}(\mathbf{z}^t(x), \mathcal{E}^*(x))$ indicate that the stationary stable state $\mathcal{E}^*(x)$ is visited at the $t$-th sample in the subsequence of plays corresponding to $\mathcal{G}(x)$. Let $\alpha_x$ denote the expected frequency of not visiting the stable state. Then, the stationary distribution of $P^{\epsilon}(x)$ is $\psi_x(\mathcal{E}^*(x))=1-\alpha_x$. We can treat $\mathbbm{1}(\mathbf{z}^t(x), \mathcal{E}^*(x))$ as a weight function of the random walk, s.t. the expected total weight is
\begin{equation}\
  \label{eq_expected_number_of_visits}
    E\left\{\frac{1}{T_1(x)}\sum_{t=1}^{T_1(x)}\mathbbm{1}(\mathbf{z}^t(x), \mathcal{E}^*(x))\right\}=1-\alpha_x,
\end{equation}
as $T_1(x)\rightarrow \infty$.

According to (\ref{eq_maximum_count}), we know that an optimal NE is guaranteed to be played during the exploration phase when the majority of trial-and-error learning plays visit $\mathcal{E}^*(x)$. Namely, $\Pr(A_x)$ is larger than the probability of the event $\sum_{t=1}^{T_1(x)}\mathbbm{1}(\mathbf{z}^t(x), \mathcal{E}^*(x))\ge T_1(x)/2$. Equivalently, we obtain
\begin{equation}
  \label{eq_single_process_bound_1}
    \Pr(\overline{A}_x) = 1-\Pr({A}_x)
    \le\Pr(\displaystyle\sum_{t=1}^{L_1(x)} \mathbbm{1}(\mathbf{z}^t(x), \mathcal{E}^*(x))\le \frac{T_1(x)}{2}).
\end{equation}
Then, following~\cite[Theorem 3.1]{chung_et_al_2912}, we have
\begin{equation}
  \label{eq_single_process_bound}
  \begin{array}{ll}
    \Pr(\overline{A}_x)
    \stackrel{\textrm(a)}{\le}\Pr(\displaystyle\sum_{t=1}^{L_1(x)} \mathbbm{1}(\mathbf{z}^t(x), \mathcal{E}^*(x))\le (1-\rho)\psi_x(\mathcal{E}^*(x))T_1(x))
    \le c_x\Vert\phi_x\Vert_{\psi_x} \exp\left(\displaystyle\frac{-\rho^2\psi_x(\mathcal{E}^*(x))T_1(x)}{72\tau_x(\frac{1}{8})} \right),
  \end{array}
\end{equation}
where (a) follows (\ref{eq_single_process_bound_1}) by setting $(1-\rho)\psi_x(\mathcal{E}^*(x))\ge\frac{1}{2}$, $\Vert\phi_x\Vert_{\psi_x}\triangleq\sqrt{\sum_{\mathbf{z}(x)\in\mathcal{Z}(x)}\frac{\psi^2_x(\mathbf{z}(x))}{\phi_x(\mathbf{z}(x))}}$, and $\tau_x(\frac{1}{8})$ is the mixing time of the Markov process $P^{\epsilon}(x)$ for an accuracy of $1/8$ (see~\cite[Theorem 3.1]{chung_et_al_2912}). By selecting a sufficiently small $\epsilon$ for each $P^{\epsilon}(x)$, we are able to adopt a unique target stable probability $\forall x\in\mathcal{X}: \psi_x(\mathcal{E}^*(x))\ge\psi$, where $\psi$ is a constant. We note that the right-hand side of (\ref{eq_single_process_bound}) is a monotonically decreasing function of $\psi_x(\mathcal{E}^*(x))$. Then, we can set $\psi_x(\mathcal{E}^*(x))=\psi$. To ensure $0<\rho<1$, from $(1-\rho)\psi\ge\frac{1}{2}$ we obtain $\rho\le1-\frac{1}{2\psi}$ and $\psi>\frac{1}{2}$. Then, we can choose $\rho=1-\frac{1}{2\psi}$ and obtain
\begin{equation}
  \label{eq_probability_bound_good_actions_2}
  \begin{array}{ll}
    \Pr(\overline{A}_x) \le c_x\Vert\phi_x\Vert_{\psi_x} \exp\left(-\displaystyle\frac{(1-\frac{1}{2\psi})^2\psi T_1(x)}{72\tau_x(\frac{1}{8})} \right).
  \end{array}
\end{equation}

We know that $(1-\frac{1}{2\psi})^2\psi$ is a monotonically increasing function if $\psi>\frac{1}{2}$. Denote $\overline{c}^{\max}_{\mathcal{X}}=\max\limits_{x\in\mathcal{X}} c_x\Vert\phi_x\Vert_{\psi_x}$, and $T_1(x)=\omega(x)T_1$, where $\sum\limits_{x\in\mathcal{X}}\omega(x)=1$. Then, for $\psi>\frac{1}{2}$, the right-hand side of (\ref{eq_probability_bound_good_actions_2}) is a monotonically decreasing function of $\psi$. Thereby, we can pick $\psi\ge\frac{2}{3}$ (consequently, $\rho\le\frac{1}{4}$), and obtain
\begin{equation}
  \label{eq_probability_bound_good_actions_3}
  \begin{array}{ll}
    \Pr(\overline{A}_x)\le c_x\Vert\phi_x\Vert_{\psi_x} \exp\left(-\displaystyle\frac{\psi-1+\frac{1}{4\psi} }{72\tau_x(\frac{1}{8})}T_1(x)\right)
    \le \overline{c}^{\max}_{\mathcal{X}} \exp\left(-\displaystyle\frac{\omega(x)T_1}{1728\tau_x(\frac{1}{8})}\right).
  \end{array}
\end{equation}
Then, by (\ref{eq_probability_bound_good_actions}) and (\ref{eq_probability_bound_good_actions_3}) we know that the error probability after running $g(k)=c_2k^{\delta}$ rounds of trial-and-error learning is
\begin{equation}
  \label{eq_probability_bound_learning_error}
  \begin{array}{ll}
    \Pr\nolimits_{l}^k\le \overline{c}^{\max}_{\mathcal{X}}\displaystyle \sum_{x\in\mathcal{X}}\exp\left(-\displaystyle\frac{\omega(x)T_1}{1728\tau_x(\frac{1}{8})}\right)
    \le X\overline{c}^{\max}_{\mathcal{X}} \displaystyle\exp\left(-\frac{\omega(\overline{x})}{1728\tau_{\overline{x}}(\frac{1}{8})}c_2k^{\delta}\right),
  \end{array}
\end{equation}
where $\overline{x}=\arg\min_{x}{\omega(x)}/{\tau_x(\frac{1}{8})}$, and by construction of Algorithm~\ref{alg_bandit} we have $T_1=c_2k^{\delta}$. Again, since the right-hand side of (\ref{eq_probability_bound_good_actions_2}) is a monotonically decreasing function, we can always find an epoch $k$ ensuring that the upper bound of $\Pr\nolimits_{l}^k$ shrinks to a sufficiently small value.

\subsection{Regret Bound of Algorithm~\ref{alg_bandit}}
\label{subsec_regret_analysis}
Now, we have the probabilities of errors propagated from the exploration phase, i.e., $\Pr_e^k$ and the learning phase i.e., $\Pr_l^k$ bounded by (\ref{eq_bound_of_exploration_probability}) and (\ref{eq_probability_bound_learning_error}), respectively. Thereby, we are ready to provide the formal proof to Theorem~\ref{thm_main} in the following discussion. We assume that the mild conditions such as the condition of discernible arm-values in Lemma~\ref{la_good_allocation} are satisfied by the considered contextual bandit game. Recall that we set in Algorithm~\ref{alg_bandit} $f(k)=c_1k$, $g(k)=c_2k^{\delta}$ and $h(k)=c_32^k$ in each epoch of the learning process. We suppose that the total number of epoch is $K$, s.t.
\begin{equation}
  \label{eq_epoch_round_relation}
    T\ge\sum_{k=1}^{K-1}(c_1k+c_2k^{\delta}+c_32^k)\ge\sum_{k=1}^{K-1}c_32^k\ge c_3(2^K-2).
\end{equation}
Then, by taking logarithm to both sides of the inequality in (\ref{eq_epoch_round_relation}), we can derive the logarithmic upper bound of $K$ as $K\le\log_2({T}/{c_3}+2)$. The total regret incurred by the learning scheme in Algorithm~\ref{alg_bandit} is composed of three parts, namely, the regret due to action exploration, trial-and-error learning and due to sub-optimal (erroneous) policies in the exploitation phases. We note that for each round of play the total regret of the $M$ players could be as large as $M$. Then, we obtain the regret bound of a single epoch in the form of (\ref{eq_bound_form}).
\begin{equation}
  \label{regret_bound_in_epoch}
  \begin{array}{ll}
    \Delta R^k\le M(c_1+c_2k^{\delta}) + M(\Pr\nolimits^k_e+\Pr\nolimits^k_l)c_32^k
    \le\! M(c_1\!+\!c_2k^{\delta}) \!+\!
    \displaystyle M\left(4ML^Xe^{-k} \!+\!  X\overline{c}^{\max}_{\mathcal{X}} e^{-\frac{\omega(\overline{x})c_2k^{\delta}}{1728\tau_{\overline{x}}(\frac{1}{8})}}\right)c_32^k.
  \end{array}
\end{equation}
We note that with $\delta>1$ there exists an epoch $k_0$, s.t. $\forall k\ge k_0, \exp\left(-\frac{\omega(\overline{x})}{1728\tau_{\overline{x}}(\frac{1}{8})}c_2k^{\delta-1}\right)\le {1/e}$. Define $A_0=4ML^X+X\overline{c}^{\max}_{\mathcal{X}}$. Then, from (\ref{regret_bound_in_epoch}) we obtain that with $\beta=\frac{2}{e}<1$, for $k>k_0$
\begin{eqnarray}
  \label{regret_bound_in_epoch_2}
    \begin{array}{ll}
      \Delta R^k \le M(c_1+c_2k^{\delta}) +M\left(4ML^X+X\overline{c}^{\max}_{\mathcal{X}}\right)c_3e^{-k}2^k
      \le M(c_1+c_2k^{\delta}) + A_0c_3M\beta^k,
    \end{array}
\end{eqnarray}
Since the second term of the right-hand side of (\ref{regret_bound_in_epoch_2}) vanishes exponentially with $k$, we obtain that for some constant $\overline{C}$ representing the constant regret until the first $k_0$ epoch,
\begin{equation}
  \label{regret_bound_in_T}
  \begin{array}{ll}
    \Delta R_T =\displaystyle\sum_{k=1}^K \Delta R^k
     \le \overline{C} + \displaystyle
    M\sum_{k=k_0+1}^{K}c_1 + c_2M\sum_{k=k_0+1}^{K}k^{\delta} + A_0c_3M\sum_{k=k_0+1}^{K}\beta^k\\
    \stackrel{\textrm{(a)}}{\le} \displaystyle\overline{C} + c_1M\log_2\left(\frac{T}{c_3}+2\right) + A_0c_3M\frac{1-\beta^{K+1}}{1-\beta} +  c_2M\log_2^{1+\delta}\left(\frac{T}{c_3}+2\right)\\
    \stackrel{\textrm{(b)}}{\le} \displaystyle\overline{C}_1 +  c_1M\log_2\left(\frac{T}{c_3}+2\right) + c_2M\log_2^{1+\delta}\left(\frac{T}{c_3}+2\right),
  \end{array}
\end{equation}
where (a) is obtained by replacing $K$ with $\log_2({T}/{c_3}+2)$, and (b) is obtained by replacing $\overline{C}$ with $\overline{C}_1=\overline{C} + \frac{A_0c_3M}{1-\beta}$. Then, (\ref{regret_bound_in_T}) completes the proof to Theorem~\ref{thm_main}.

\begin{Remark}[Computational Complexity of Algorithm~\ref{alg_bandit}]
With the regret bound after $T$ rounds of plays given by (\ref{regret_bound_in_T}), now we examine the computational complexity of Algorithm~\ref{alg_bandit} in a single time slot $t$. We note that the one-step action perturbation (Line 1) and auxiliary state transition (Lines 2-21) in Algorithm~\ref{alg_trial_and_error} are completed in constant time. Then, it suffices to compare the computational complexity of one single round in the exploration phase, the trial-and-error phase and the exploitation phase, respectively. Again, the expected arm-value estimator is updated in constant time in a given time slot of the exploration phase (see (\ref{eq_update_mean}) or Lines 5-8 of Algorithm~\ref{alg_bandit}). For a time slot in the trial-and-error phase, it may take $O(XL)$ for a link to construct the intermediate game and the corresponding auxiliary states (see Lines 11-16 of Algorithm~\ref{alg_bandit}) at the beginning of the phase. Comparatively, for the exploitation phase, finding the maximum number of auxiliary-state visits takes a player $O(L)$ in one time slot (see (\ref{eq_maximum_count})).

Additionally, the space complexity of Algorithm~\ref{alg_bandit} in one single round is $O(XL)$ for the arm-value estimator record-keeping (see (\ref{eq_update_mean}) with respect to different contexts) in the exploration phase and $O(XL)$ for record keeping with respect to the frequency of visiting the content states in the trial-and-error phase and the exploitation phase (see (\ref{eq_frequency_visit})). With such complexity levels, the proposed algorithm is suitable for resource-limited IoT devices.
\end{Remark}

\section{Adaptation to Unobservable Contexts}
\label{sec_unobservable}
Now, we consider that the contexts are no longer released/observable at the beginning of each time slot. This is corresponding with the situation when no energy detector or primary base station feedback is available to the IoT devices. Since the
intermediate game $\mathcal{G}(x)$ cannot be established for discernible context $x$ in this situation, policy learning in the trial-and-error phase is reduced to one single perturbed Markov chain. Recall that the joint distribution of the state and the values of each arm follows $(x, r_m)\sim D_m$. Therefore, learning arm-selection without discerning the underlying context $x$ is reduced to a normal MP-MAB, where the value distribution of each arm follows the marginal distribution over $r_m\sim D_{m,r_m}$: $p_m(r_m)=E_x\{p_{r_m|x}(r_m|x)\}$. Without making any significant change to the proposed algorithm, we define a modified regret in Definition~\ref{def_regret_2} (cf., Definition~\ref{def_regret}):
\begin{Definition}[Modified Regret]
  \label{def_regret_2}
  Let the expected reward of a policy $\pi$ without discerning $x$ be denoted by $V(\pi)=E_{\mathbf{r}^t\sim D_{m,r_m}}\left\{\sum_{m=1}^M {v}_m^t(\pi)\right\}$. For the series $\{\mathbf{r}^t\}^T_{t=1}$,  drawn from the distribution $D_{m,r_m}$,
 the expected regret of algorithm $\mathcal{B}=\{\tilde{\pi}_1,\ldots,\tilde{\pi}_T\}$ with respect to a policy $\pi$ is
  \begin{equation}
    \label{label_regret_wrt_pi_2}
    \Delta R(\mathcal{B}, \pi, T) = TV(\pi) - E\left[\sum_{t=1}^T\sum_{m=1}^M v^t_{m}(\tilde\pi^t_{m})\right].
  \end{equation}
  The regret of algorithm $\mathcal{B}$ with respect to policy space $\Pi$ is
  \begin{equation}
    \label{label_regret_2}
    \Delta R(\mathcal{B}, \Pi, T) = \sup_{\pi\in\Pi} TV(\pi) - E\left[\sum_{t=1}^T\sum_{m=1}^M v^t_{m}(\tilde\pi^t_{m})\right].
  \end{equation}
\end{Definition}

Compared with Definition~\ref{def_regret}, the regret defined in Definition~\ref{def_regret_2} will lead to a sub-optimal allocation solution because (\ref{label_regret_2}) requires algorithm $\mathcal{B}$ to produce a consistent joint policy for all different context $x$. Based on the modified regret in Definition~\ref{def_regret_2}, for an arbitrary, individual policy $\pi_m$ that results in no collision, the true expected reward becomes $E_{(r_m)\sim D_{m,r_m}}\{r_{m, {\pi}_m}\}$. By the law of unconscious statistician we have $E_{(r_m)\sim D_{m,r_m}}\{r_{m, {\pi}_m}\}=E_{(x, r_m)\sim D_m}\{r_{m, {\pi}_m}\}$. Following our discussion in Section~\ref{sub_sec_prob_exploration}, the unbiased reward estimator based on $\mathcal{W}_m$ is now $\hat{\mu}_m(\pi_m)=\frac{1}{N_m}\sum_{i=1}^{N_m} \frac{\mathbbm{1}(\pi_m, a^i_{m})v^i_{m}}{1/L}$. Then, (\ref{eq_var_bound}) still holds for the variation of the new estimator and for (\ref{eq_estimation_bound}) we have a slightly different bound
  \begin{equation}
    \label{eq_estimation_bound_unobservable}
    \begin{array}{ll}
      \Pr\left( \sup\limits_{m\in{\mathcal{M}}, \pi_m}\left\{\left(\hat{\mu}_{m}(\pi_m)\!-\!E\{r_{m,\pi_m}\}\right)\!>\!\eta \Big\vert \forall m: \vert W_m\vert \!\ge\! C \right\}\right)
      \le 2ML\exp\left(-\displaystyle\frac{C\eta^2}{2L+2c\eta/3}\right),
    \end{array}
  \end{equation}
since the learning algorithm no longer discerns the underlying context $x$. Subsequently, for an exploration phase that lasts for $c_1k$ rounds in epoch $k$, we have a new probability bound for the exploration error:
\begin{equation}
  \label{eq_bound_of_exploration_probability_unobservable}
  \begin{array}{ll}
    \displaystyle\Pr\nolimits^k_e=\gamma\le 4MLe^{-k}.
  \end{array}
\end{equation}
Then, the inequality in Lemma~\ref{la_lamma_exploreation_error} becomes $T_0\ge \max\left(16L\frac{L+c\eta/3}{\eta^2}\ln\left(\frac{4ML}{\gamma}\right), 32L\ln\left(\frac{2M}{\gamma}\right)\right)$.
Therefore, keeping the same value of $f(k)=c_1$ for the exploration phase still guarantee the accuracy of arm-value estimator under the marginal distribution. This leads to the modified exploration phase in Algorithm~\ref{alg_exploration_unobservable}.
\begin{algorithm}[!t]
    \begin{small}
 \caption{Modified exploration phase for player $m$ with non-observable context at the $k$-th epoch.}
 \begin{algorithmic}[1]
  \FOR {$t = 1, \ldots, f(k)$}
  \STATE Sample an arm $a^t_{m}\in\{1,\ldots,L\}$ uniformly at random and observe the feedback $(a^t_{m}, v^t_{m}(\mathbf{a}^t))$
  \IF {$v^t_{m}(\mathbf{a}^t)\ne0$}
    \STATE $\mathcal{W}_m\leftarrow \mathcal{W}_m\cup \{(a^t_{m}, v^t_{m}(\mathbf{a}^t))\}$
    \STATE Estimate the expected value of arm $l=a^t_{m}$: $\mu^k_{m,l}\leftarrow \frac{\sum\limits_{(a_{m}, v_{m})\in\mathcal{W}_m}v^t_{m}\mathbbm{1}(a_m, l)} {\sum\limits_{(a_{m}, v_{m})\in\mathcal{W}_m}\mathbbm{1}(a_m, l)}$
  \ENDIF
  \ENDFOR
 \end{algorithmic}
 \label{alg_exploration_unobservable}
\end{small}
\end{algorithm}

Thanks to the phase-based structure of the learning scheme, we are able to isolate the trial-and-error phase from the exploration phase for arm-value estimation. When the arm-values of the unique intermediate game $\mathcal{G}$ for each epoch is provided by Algorithm~\ref{alg_exploration_unobservable}, it is straightforward to prove that Theorem~\ref{th_stable_states} and Proposition~\ref{prop_unique_NE} still hold with Algorithm~\ref{alg_trial_and_error}. The only difference lies in that the multiple sub-events $A_x$ in (\ref{eq_probability_bound_good_actions}) is now replaced by a single event $A$. Therefore, without any modification to the discussion of regret bound in Section~\ref{subsec_regret_analysis}, we can show that Theorem~\ref{thm_main} still holds with exactly the same form of bound:
\begin{equation}
    \label{eq_bound_regret_2}
    \begin{array}{ll}
        \Delta R_T \le \displaystyle\overline{C}_1 +  c_1M\log_2\left(\frac{T}{c_3}+2\right) + c_2M\log_2^{1+\delta}\left(\frac{T}{c_3}+2\right)
        =O(M\log_2^{1+\delta}(T)).
    \end{array}
\end{equation}

\section{Simulation Results}
\label{sec_simulation}
\begin{table*}[!t]
  \centering
  \caption{Main Parameters of Algorithm~\ref{alg_bandit} Used in the Experiments}\vspace*{-1mm}
  \scriptsize
 \begin{tabular}{|c c | c c | c c | c c | c c | c c|}
 \hline
 Parameter & Value & Parameter & Value & Parameter & Value & Parameter & Value & Parameter & Value & Parameter & Value\\
 \hline
 $\epsilon$ & $0.01$ & $\xi$ & $0.001$ & $\delta$ & 1 & $c_1$ & 100 & $c_2$  & $200$ & $c_3$ & $100$\\
 \hline
 $\alpha_{11}$ & $-0.12$ & $\alpha_{12}$ & $0.15$ & $\alpha_{21}$ & $-0.35$ & $\alpha_{22}$ & $0.4$ & Reward range & $[0,1]$ &  & \\ \hline
\end{tabular}
\label{table_setting}
\end{table*}

\subsection{Evaluation of the Proposed Algorithm}
\label{Subsec:AlgEvaluation}
We first demonstrate the efficiency of Algorithm~\ref{alg_bandit} using a toy-like example (for convenience of illustration) of a contextual MP-MAB setup of 2 players, 3 arms and 3 contexts, where for each player, the contexts and arm values are jointly sampled from discrete uniform distributions. For comparison, we implement two categories of non-contextual algorithms, the ``Musical Chairs'' (MC) algorithm~\cite{rosenski2016multi} and its variation ``SIC-MMAB'' (i.e., Synchronisation Involves Communication in Multiplayer Multi-Armed Bandits) algorithms~\cite{boursier:hal-02371008} and another three-phase-epoch-based decentralized learning algorithm, the Game of Thrones (GoT) algorithm~\cite{bistritz2018distributed}. We also adopt the Hungarian algorithm to indicate the ground-truth socially optimal arm-allocation with a centralized allocator\footnote{The source code and configurations of our experiments can be found at \url{https://github.com/wbwang2020/MP-MAB}.}. In the first experiments, we adopt the main parameters for the trial-and-error learning algorithm as Table I.
\begin{figure}[!t]
\centering     
\subfigure[]{\label{fig_efficiency_deomonstration_a}\includegraphics[width=.32\textwidth]{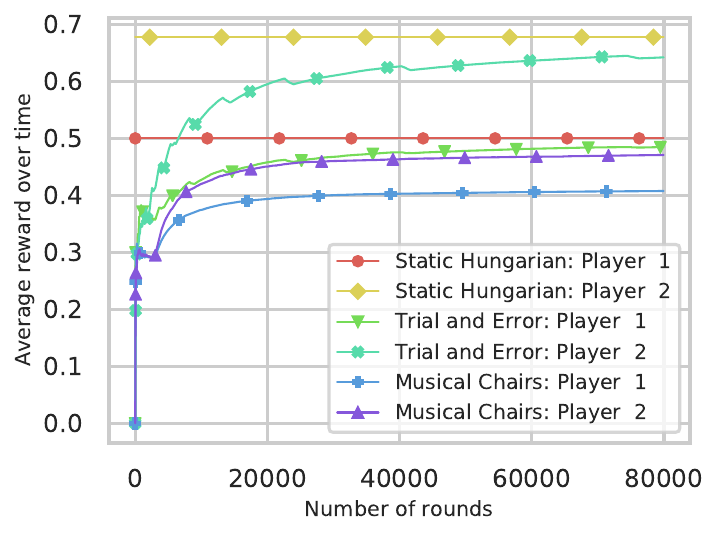}}
\subfigure[]{\label{fig_efficiency_deomonstration_b}\includegraphics[width=.32\textwidth]{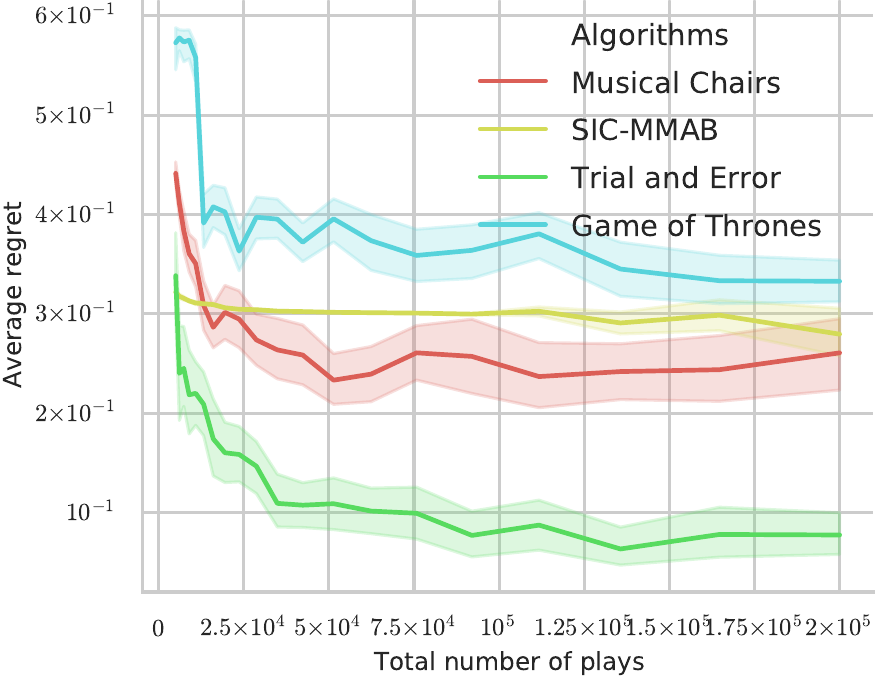}}
\subfigure[]{\label{fig_regret_evolution}\includegraphics[width=.32\textwidth]{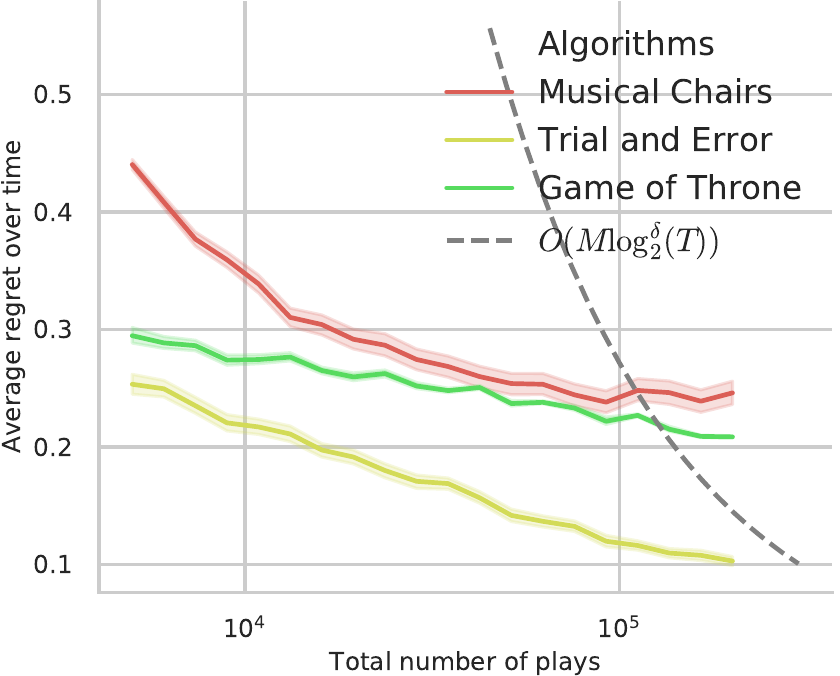}}
\vspace{-2mm}
\caption{(a) Average payoff of individual players with respect to time. (b) Average regrets of plays vs. different time horizons $T$. (c) Regret evolution with respect to the total number of plays $T$.}
\label{fig_efficiency_deomonstration}
\end{figure}

Figure~\ref{fig_efficiency_deomonstration_a} provides an intuitive illustration on the evolution of the players' average rewards as the bandit algorithms progress over time. In Figure~\ref{fig_efficiency_deomonstration_a}, the curves marked as ``Static Hungarian'' indicate the expected rewards of each individual players (Figure 2(a)) when the true social-optimal allocation is adopted. Figure~\ref{fig_efficiency_deomonstration_b} shows the comparison of the average regret over different time horizon for the proposed trial-and-error learning algorithm, the GoT algorithm, the MC algorithm and the SIC-MMAB algorithm. In this experiment, for the same sequence of context-arm-value sampling with a total rounds of $T$, the bandit game is repeatedly played for 200 times with each algorithm. The shaded areas around the solid curves indicates the empirical performance variation during the Monte Carlo simulations. The gap between the regrets of trial-and-error learning and the other algorithms clearly indicates that the proposed algorithm is able to better adapt to the stochastic evolution of the contextual dimension.

In Figure~\ref{fig_regret_evolution}, we compare the evolution of the average regrets of trial-and-error learning, MC and the GoT algorithm~\cite{bistritz2018distributed} in a slightly larger problem with 5 arms, as the total number of plays (horizons) $T$ increases. The dashed curve ``$O(M\log_2^\delta(T))$'' represents a heuristic regret bound in the same form as (\ref{regret_bound_in_T}), where the heuristic bound has a set of parameters as $c_1M=200$, $c_2M=40$ and $\overline{C}_1=0$ for the considered game. Apart from the similar finding that the two learning-based algorithms achieves a better performance than MC, it is worth noting from the simulation result that the proposed trial-and-error learning algorithm has a much faster convergence rate than GoT.

\subsection{Algorithm Evaluation in Heterogeneous IoT over Shared Bandwidth}
\label{Subsec:HetNet_Experiment}
Now, we apply the proposed channel allocation algorithm to the simulated scenario of an ad-hoc IoT network underlaying over the spectrum bands licensed to a cellular primary network. We perform a series of simulations with the focus on the following measurements: (a) the sum of the normalized link data rates (i.e., rewards measured in achievable throughputs), (b) the frequency of collision and channel-switching during policy learning and (c) the scalability of the proposed algorithm. Throughout the simulations, we consider that the channel statistics are unknown and heterogeneous with respect to different IoT links. We also consider that different levels of interference from the primary transmission is caused by a number of primary users randomly occupying and leaving the spectrum. Correspondingly, the contexts of the mapped MP-MAB reflect the IDs and the power levels of different licensed users. We consider that a fixed number of randomly distributed IoT devices move in random and slow motion (e.g., following a Gauss-Markov mobility model~\cite{doi:10.1002/wcm.72}) and underlay over the frequency of the primary users for their own transmission. For the IoT network, the entire spectrum is divided into a fixed number of logical channels\footnote{For instance, in NB-IoT-like networks, this could be implemented by grouping the OFDM symbols into a fixed number of $L$ available resource blocks. In each physical resource block a device experiences independent path loss and shadow fading, but faces the stochastic interference of the same transmit-power level from the underlying transmission of the UEs in the macrocell.}.

\begin{figure*}[!t]
\centering     
\subfigure[]{\label{fig_monte_carlo_rewards}\includegraphics[width=.31\textwidth]{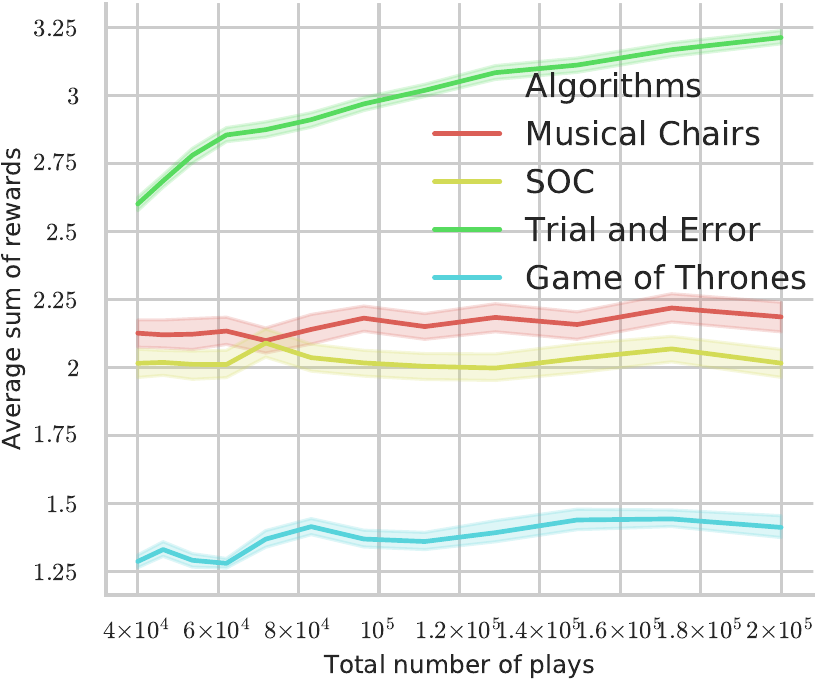}}
\subfigure[]{\label{fig_monte_carlo_collision}\includegraphics[width=.31\textwidth]{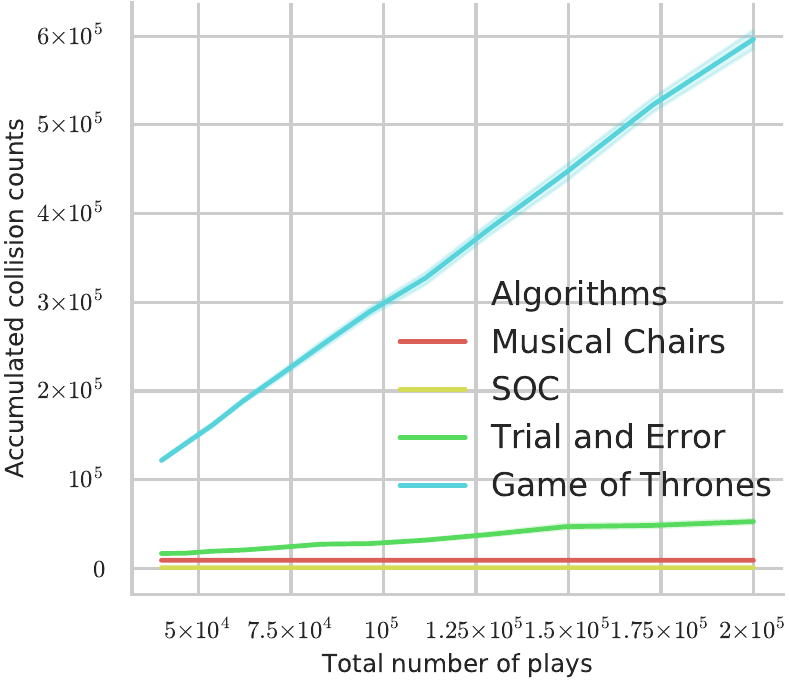}}
\subfigure[]{\label{fig_monte_carlo_switching}\includegraphics[width=.31\textwidth]{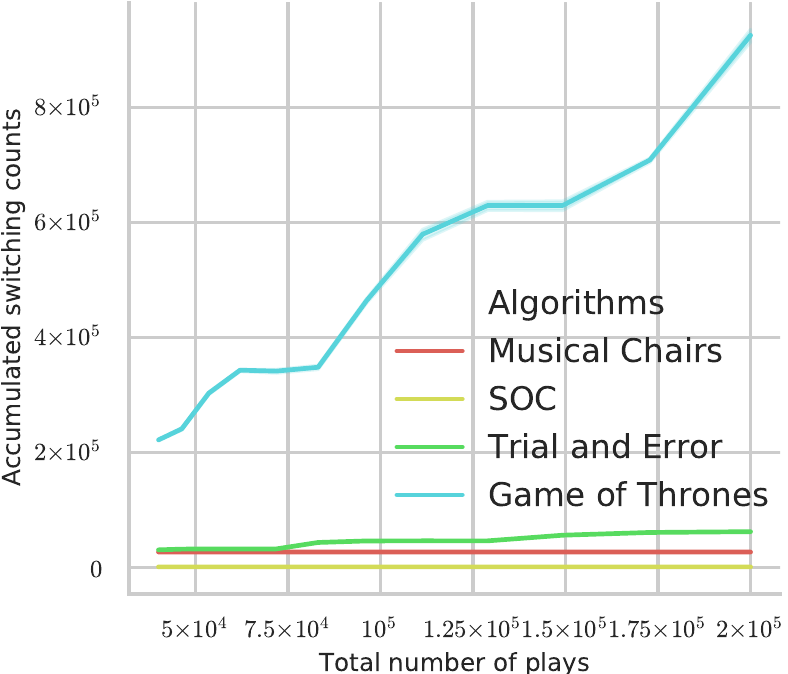}}\vspace{-2mm}
\caption{(a) Average sum of normalized rates over time with respect to operational horizons. (b) Accumulated collisions among IoT devices with respect to time horizons. (c) Accumulated channel switching counts among IoT devices with respect to time horizons.}
\label{fig_10_device_result}
\end{figure*}

In Figure~\ref{fig_10_device_result}, we demonstrate the simulation results for a setting of 10 IoT devices over 12 logical channels of the bandwidth, which is randomly accessed by 3 licensed users of 2 power levels. In addition to the two categories of reference MP-MAB algorithms, i.e., MC and GoT, we also compare the proposed algorithm with another
state-of-the-art MP-MAB algorithm based on channel swapping, i.e., ``Stable Orthogonal Channel (SOC)'' allocation~\cite{8792108}. Compared with MC, SOC is able to address the heterogeneous distribution of arm-values in a non-contextual setting, while it aims to achieve stable non-colliding allocation instead of social-optimal network performance as with GoT. Figure~\ref{fig_monte_carlo_rewards} clearly shows that our proposed scheme achieves the best performance out of the 4 algorithms. Figure~\ref{fig_monte_carlo_collision} indicates that such better performance of the proposed algorithm is achieved at the cost of slightly more collisions, due to more frequent policy explorations with respect to the contexts over time.

Furthermore, it is worth noting from Figure 3(c) that by considering the influence of contexts, trial-and-error learning experiences more frequent channel switching than the non-contextual algorithms (i.e., MC and SOC). More specifically, the switching frequency measures the consistency of action-taking against different context by different algorithms. The lower the frequency, the higher the policy consistency is across different contexts. This is in accordance with our discussion in Section~\ref{sec_unobservable}, that the contextual MP-MAB algorithm is able to provide more flexible policies to gain better network performance than the normal MP-MAB algorithms, which blindly choose the same policy for different contexts. However, we note that GoT has significantly higher collision frequency and channel switching frequency. This indicates that trial-and-error learning has a significantly higher convergence rate than GoT, even when the size of the auxiliary-state transition graph (see also our discussion in Section~\ref{subsec_trial_and_error}) for policy learning of GoT is smaller than our proposed algorithm due to ignoring the contexts. We believe this is the main reason for GoT to experience excessive collisions in Figure~\ref{fig_monte_carlo_collision} during the same time horizon $T$. In other words, with the existence of contexts, GoT needs a much longer policy-learning phase (equivalently, a larger $c_2$) than trial-and-error
learning to achieve a better performance than MC.

\begin{figure*}[!t]
\centering     
\subfigure[]{\label{fig_network_rewards}\includegraphics[width=.31\textwidth]{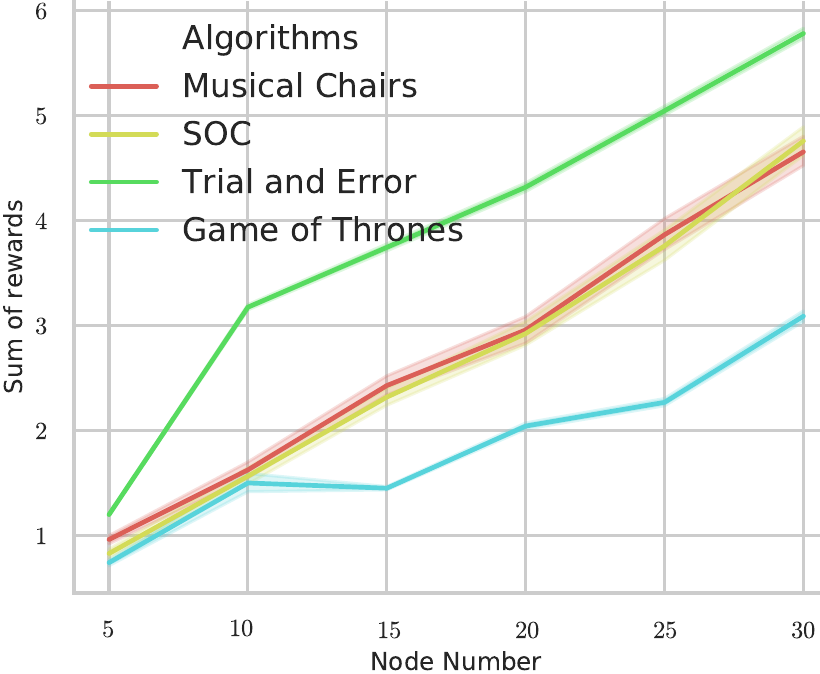}}
\subfigure[]{\label{fig_network_collision}\includegraphics[width=.31\textwidth]{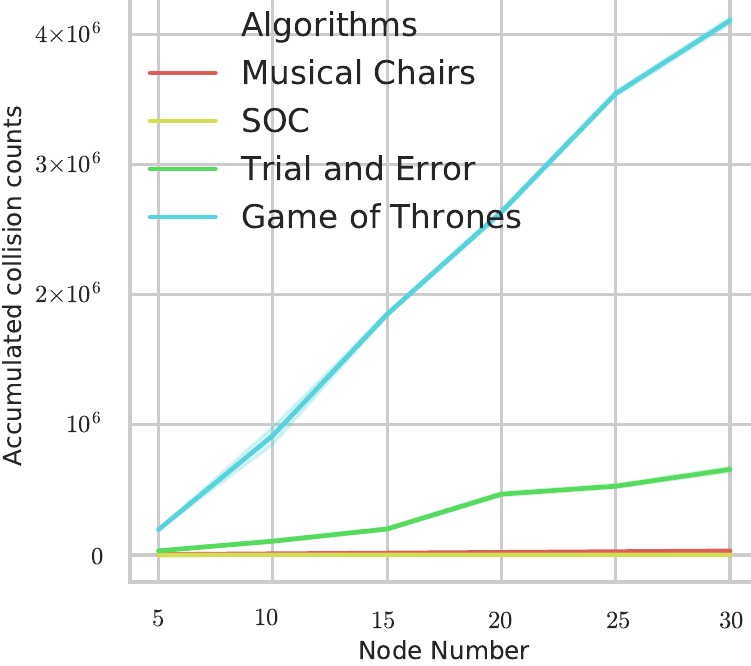}}
\subfigure[]{\label{fig_network_switching}\includegraphics[width=.31\textwidth]{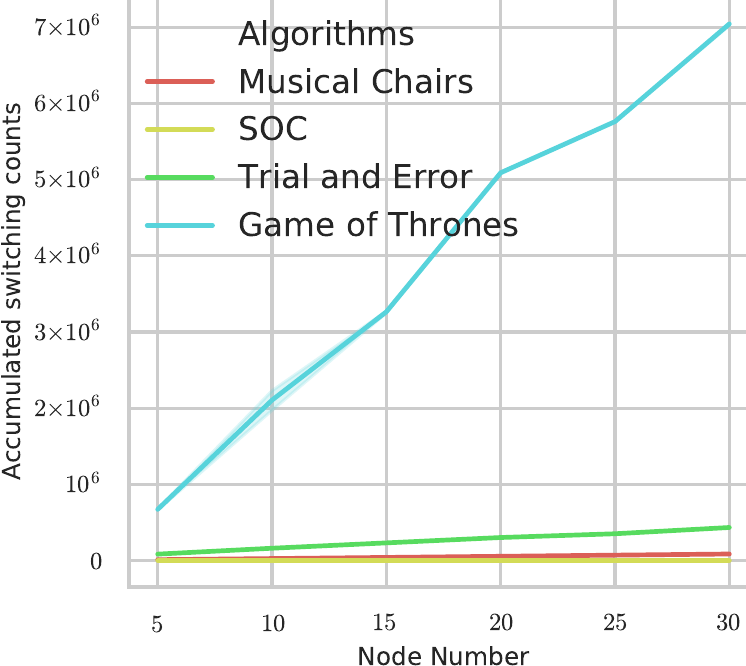}}\vspace{-2mm}
\caption{(a) Average sum of normalized rates over time with respect to network size. (b) Accumulated collisions among IoT devices with respect to network size. (c) Accumulated channel switching counts among IoT devices with respect to network size.}
\label{fig_network_result}
\end{figure*}

Finally, we examine the scalability of different algorithms with respect to the network size in Figure~\ref{fig_network_result}, for which we fix the horizon of simulations to be $4\times10^5$ rounds for different network sizes varying from 5 links to 30 links. For for each epoch in trial-and-error learning and GoT, the length of the
perturbation-based learning phase is set to start with $c_2=3\times10^3$ for a network of 5 nodes and then increase proportionally as the network size grows. As shown in Figures~\ref{fig_network_collision} and~\ref{fig_network_switching}, the proposed trial-and-error learning algorithm and GoT experience more collisions than MC and SOC, as the network size increases. This is mainly due to both the significantly longer policy exploration and the larger auxiliary state space as the network size increases. Again, the GoT algorithm needs significantly larger number of rounds to achieve the same level of performance as MC and SOC when the network size increases. This is mainly due to both the significantly longer policy exploration (i.e., controlled by parameter $f(k) = c_1$) and the larger auxiliary state space (i.e., controlled by parameter $g(k) = c_2k^{\delta}$) as the network
size increases. Again, GoT needs significantly larger number of rounds to achieve the same level of performance as MC and SOC when the network size increases.
As a result, it may not scale well with the network size. Comparatively, our proposed algorithm is able to achieve the better performance (see Figure~\ref{fig_network_rewards}) than the other reference algorithms at an acceptable cost of more frequent collisions (see Figure~\ref{fig_network_collision}) due to a longer learning phase.

\section{Conclusion}\label{Sec:Conclusion}
In this paper, we have proposed a decentralized, epoch-based channel-allocation algorithm based on trial-and-error learning for IoT networks underlaying over the bandwidth shared by primary users. The proposed algorithm exploits the information of primary transmissions in a framework of contexutal of multi-player, multi-armed bandits. It guarantees socially-optimal performance through repeatedly constructing intermediate non-cooperative games for performing decentralized policy learning between the phases of channel-quality exploration and policy exploitation. The proposed algorithm efficiently addresses the situation of time-varying channels with underlying unpredictable interference from the licensed transmissions. Theoretical analysis proves that the proposed policy-learning scheme is able to achieve the optimal regret in $O(M\log_2^{1+\delta}T)$ ($\delta>0$) for a contextual multi-player bandit game of $M$ players along a time horizon of $T$. Our proposed algorithm is especially appropriate for deployment in infrastructure-less networks with rigid constraint on communications between links. Particularly, the only information needed by the algorithm is the inter-link collision states over channels from the receiving device's feedback. The simulation results show that the proposed algorithm is able to achieve better performance than a number of state-of-the-art reference schemes when the IoT network underlays on realistic channels of a licensed cellular network.

\linespread{1.23}
\bibliographystyle{IEEEtran}
\bibliography{bibfile}

\end{document}